\documentclass{ieeetj}
\usepackage{cite}
\usepackage{amsmath,amssymb,amsfonts}
\usepackage{algorithmic}
\usepackage{graphicx,color}
\usepackage{textcomp}
\usepackage{xcolor}
\usepackage{hyperref}
\usepackage{multirow}
\hypersetup{hidelinks=true}
\usepackage{algorithm,algorithmic}
\def\BibTeX{{\rm B\kern-.05em{\sc i\kern-.025em b}\kern-.08em
    T\kern-.1667em\lower.7ex\hbox{E}\kern-.125emX}}
\AtBeginDocument{\definecolor{tmlcncolor}{cmyk}{0.93,0.59,0.15,0.02}\definecolor{NavyBlue}{RGB}{0,86,125}}

\def\OJlogo{\vspace{-4pt}$<$Society logo(s) and publication title will appear here.$>$}
\def\seclogo{\vspace{10pt}$<$Society logo(s) and publication title will appear here.$>$}

\def\authorrefmark#1{\ensuremath{^{\textbf{#1}}}}

\begin{document}
\receiveddate{XX Month, XXXX}
\reviseddate{XX Month, XXXX}
\accepteddate{XX Month, XXXX}
\publisheddate{XX Month, XXXX}
\currentdate{XX Month, XXXX}
\doiinfo{XXXX.2022.1234567}

\markboth{}{Khan \textit{et al.}: Blind Ultrasound Image Enhancement via Self-Supervised Physics-Guided Degradation Modeling}

\title{Blind Ultrasound Image Enhancement via Self-Supervised Physics-Guided Degradation Modeling}

\author{%
Shujaat Khan\authorrefmark{1,2},%
\;Syed Muhammad Atif\authorrefmark{3},%
\;Jaeyoung Huh\authorrefmark{4},%
\;Syed Saad Azhar Ali\authorrefmark{5,6}%
}

\affil{\authorrefmark{1}\,Department of Computer Engineering, College of Computing and Mathematics, King Fahd University of Petroleum \& Minerals (KFUPM), Dhahran 31261, Saudi Arabia}
\affil{\authorrefmark{2}\,SDAIA--KFUPM Joint Research Center for Artificial Intelligence (JRC-AI), King Fahd University of Petroleum \& Minerals, Dhahran 31261, Saudi Arabia}
\affil{\authorrefmark{3}\,Department of Computer Science and Information Technology, Sir Syed University of Engineering and Technology (SSUET), Karachi 75300, Pakistan}
\affil{\authorrefmark{4}\,Digital Technology \& Innovation Center, Siemens Healthineers, Princeton, NJ 08540, USA}
\affil{\authorrefmark{5}\,Department of Aerospace Engineering, King Fahd University of Petroleum \& Minerals (KFUPM), Dhahran 31261, Saudi Arabia}
\affil{\authorrefmark{6}\,Interdisciplinary Research Center for Smart Mobility and Logistics, King Fahd University of Petroleum \& Minerals (KFUPM), Dhahran 31261, Saudi Arabia}

\corresp{Corresponding author: Shujaat Khan (e-mail: \href{mailto:shujaat.khan@kfupm.edu.sa}{shujaat.khan@kfUPM.edu.sa}).}

\begin{abstract}
Ultrasound (US) interpretation is hampered by multiplicative speckle, acquisition blur from the point-spread function (PSF), and scanner- and operator-dependent artifacts. Supervised enhancement methods assume access to clean targets or known degradations; conditions rarely met in practice. We present a blind, self-supervised enhancement framework that jointly deconvolves and denoises B-mode images using a Swin Convolutional U-Net trained with a \emph{physics-guided} degradation model. From each training frame, we extract rotated/cropped patches and synthesize inputs by (i) convolving with a Gaussian PSF surrogate and (ii) injecting noise via either spatial additive Gaussian noise or complex Fourier-domain perturbations that emulate phase/magnitude distortions. For US scans, clean-like targets are obtained via non-local low-rank (NLLR) denoising, removing the need for ground truth; for natural images, the originals serve as targets.
Trained and validated on UDIAT~B, JNU-IFM, and XPIE Set-P, and evaluated additionally on a 700-image PSFHS test set, the method achieves the highest PSNR/SSIM across Gaussian and speckle noise levels, with margins that widen under stronger corruption. Relative to MSANN, Restormer, and DnCNN, it typically preserves an extra $\sim$1--4\,dB PSNR and 0.05--0.15 SSIM in heavy Gaussian noise, and $\sim$2--5\,dB PSNR and 0.05--0.20 SSIM under severe speckle. Controlled PSF studies show reduced FWHM and higher peak gradients—evidence of resolution recovery without edge erosion. Used as a plug-and-play preprocessor, it consistently boosts Dice for fetal head and pubic symphysis segmentation. Overall, the approach offers a practical, assumption-light path to robust US enhancement that generalizes across datasets, scanners, and degradation types.
\end{abstract}

\begin{IEEEkeywords}
Ultrasound Imaging, Speckle Denoising, Blind Deconvolution, Self-Supervised Learning, Swin UNet, Image Enhancement, Segmentation
\end{IEEEkeywords}


\maketitle

\section{INTRODUCTION}
\IEEEPARstart{U}{ltrasound} (US) imaging is one of the most widely used diagnostic tools due to its real-time capability, safety, and low cost. Nevertheless, the diagnostic value of ultrasound is often limited by speckle noise and blurring caused by the system’s point spread function (PSF). These degradations obscure subtle anatomical features and degrade the performance of downstream tasks such as segmentation and classification. Conventional image enhancement methods require explicit knowledge of noise variance or PSF characteristics, which are rarely available in practice. Recent deep learning–based approaches have achieved significant progress, but most rely on supervised training with clean ground-truth images—an unrealistic assumption in clinical settings.

Classical approaches include anisotropic diffusion, wavelet shrinkage, and non-local means (NLM). Variants such as speckle reducing anisotropic diffusion (SRAD) \cite{yu2002srad}, fractional-order diffusion \cite{bai2007fractional}, and adaptive nonlinear diffusion \cite{kollem2024fast} improve speckle suppression while preserving edges. Wavelet-based thresholding methods \cite{chang2000adaptive,liu2017efficient,onur2022improved} exploit multiscale representations for denoising, whereas NLM \cite{buades2005nonlocalmeans} and its extensions \cite{kervrann2007bayesian-nonlocalmeans,maleki2013anisotropic-nonlocalmeans,zhang2022two,kong2024improved} leverage patch similarity for noise reduction. Low-rank and sparse priors have also been explored, such as high-order SVD \cite{rajwade2012image} and non-convex optimization strategies \cite{zhang2024fast}.

Deep learning has enabled end-to-end ultrasound enhancement frameworks. CNN-based models such as DnCNN \cite{zhang2017beyond} and UNet \cite{ronneberger2015unet} are commonly adapted for medical image restoration, while autoencoder-based methods \cite{gondara2016medical} and residual dense block architectures \cite{nanthini2024dl} aim to preserve diagnostically critical details. Self-supervised techniques including Noise2Void \cite{krull2019noise2void} and Noise2Self \cite{batson2019noise2self} have demonstrated that models can be trained directly on corrupted images without clean references, although their application to ultrasound remains limited. Another key challenge is resolution loss, which is compounded by denoising. Prior work has explored adaptive beamforming \cite{khan2021switchable,khan2019deep,khan2020adaptive}, deconvolution \cite{khan2020unsupervised,khan2020unsupervised}, and physics-guided deep learning \cite{khan2021variational}, however the methods are either assumption-heavy or require access to channel data which is not available in clinical settings.

Despite these advances, three challenges remain: (i) reliance on clean ground-truth data, (ii) limited robustness to diverse degradations, and (iii) resolution loss during denoising. To address these issues, we propose a blind, self-supervised framework for ultrasound enhancement. The method leverages a Swin Convolutional UNet (SC-UNET)\cite{zhang2023practical} trained on synthetic input–target pairs derived from real ultrasound data using physics-guided degradation modeling. The augmentation pipeline introduces diverse degradations, including Fourier-domain perturbations, Gaussian and multiplicative speckle noise, and random PSF blurs, applied to both natural images and different modalities of actual ultrasound scans. For natural images, the original non-degraded images serve as ground truth, whereas for ultrasound scans, clean-like targets are approximated using non-local low-rank (NLLR) denoising \cite{zhu2017non}. This strategy eliminates the need for manually curated ground truth while exposing the model to realistic noise conditions. Furthermore, the hierarchical Swin-based encoder–decoder structure enhances both local and global feature modeling, resulting in improved speckle suppression, boundary preservation, and robustness across heterogeneous datasets.

Furthermore, recent efforts in cross-device harmonization \cite{seoni2024all} emphasize the importance of generalization across diverse ultrasound scanners. Our framework complements such efforts by providing a plug-and-play module that enhances image quality consistently, thereby improving visualization and the reliability of downstream analysis in clinical workflows.

\section{PROPOSED METHOD}
\begin{figure*}[!htbp]
    \centering
    \includegraphics[width=\linewidth]{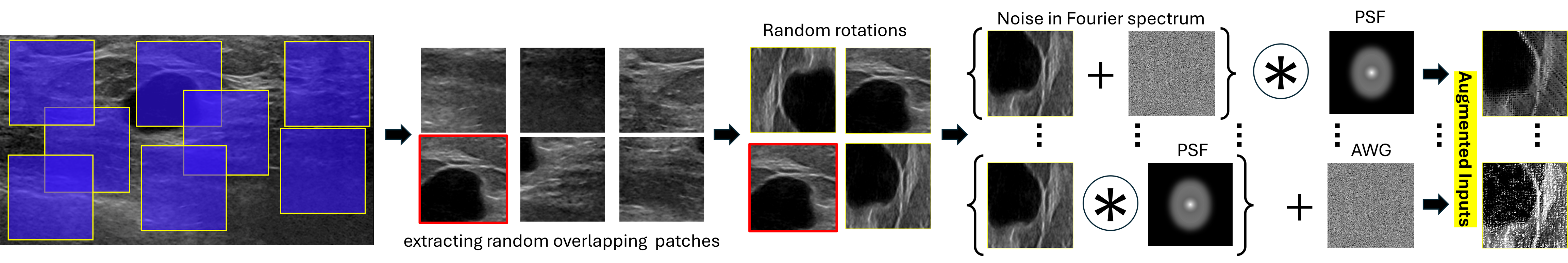}
    \caption{Physics-guided noise degradation pipeline used for training data augmentation.}
    \label{fig:physics_informed_augmentation}
\end{figure*}
\subsection{Overview}
The proposed framework is a Swin convolution-based encoder–decoder network designed to enhance grayscale ultrasound images. Training pairs are synthetically generated from real ultrasound scans using a physics-guided degradation model (shown in Fig.~\ref{fig:physics_informed_augmentation}). This model introduces additive Gaussian noise, Fourier-domain perturbations, and random blurring (to simulate PSF effects). For clean-like targets, non-local low-rank (NLLR)\cite{zhu2017non} denoising is applied, enabling self-supervised learning without requiring ground-truth clean images.

\subsection{Physics\textendash Guided Degradation Model}
\label{subsec:physics_guided_degradation}
Let $\mathbf{I}\in[0,1]^{H\times W}$ denote a grayscale ultrasound frame. During training we form patches by
(a) resizing to $128{\times}128$, (b) applying a random in\textendash plane rotation $\theta\!\sim\!\mathcal{U}[-15^\circ,15^\circ]$, and
(c) taking a random $64{\times}64$ crop of the rotated image. Denote this augmented patch by $\tilde{\mathbf{I}}$.
We then synthesize degraded inputs via two families of operators that reflect common ultrasound artifacts:

\paragraph{(1) Blur by a PSF surrogate.}
We convolve $\tilde{\mathbf{I}}$ with an isotropic Gaussian kernel of odd size
$k\!\in\!\{3,5,7,9,11,13,15,17\}$:
\begin{equation}
  \mathbf{I}_{b} = \tilde{\mathbf{I}} * h_k, \qquad
  h_k(u,v) \propto \exp\!\left(-\frac{u^2+v^2}{2\sigma_b^2}\right),
\end{equation}
with $\sigma_b$ chosen so that $k\approx 2\lceil 3\sigma_b\rceil + 1$. This approximates acquisition blur from the system PSF.

\paragraph{(2) Noise processes.}
We inject two complementary noise types:
\begin{itemize}
  \item \textbf{Spatial additive Gaussian noise}
  (to emulate thermal/receiver noise):
  \begin{equation}
    \mathbf{I}_{g} = \tilde{\mathbf{I}} + \mathbf{n}, \qquad
    \mathbf{n}\overset{\text{i.i.d.}}{\sim}\mathcal{N}(0,\sigma_g^2), \;\;
    \sigma_g\sim\mathcal{U}(0.05,0.20),
  \end{equation}
  followed by clipping to $[0,1]$.

  \item \textbf{Fourier\textendash domain complex perturbations}
  (to mimic phase/magnitude distortions and speckle\textendash like granular texture):
  \begin{equation}
    \mathcal{F}(\mathbf{I}_{f})
      = (1-\gamma_f)\,\mathcal{F}(\tilde{\mathbf{I}})
        + \gamma_f\,\|\mathcal{F}(\tilde{\mathbf{I}})\|_{\infty}\,\boldsymbol{\zeta}, \quad
    \boldsymbol{\zeta}\sim\mathcal{CN}(0,\mathbf{I}),
  \end{equation}
  where $\mathcal{CN}$ denotes zero\textendash mean complex Gaussian noise (independent real/imaginary parts),
  $\gamma_f\!\sim\!\mathcal{U}(0,0.2)$ in training, and $\mathbf{I}_{f}=\big|\mathcal{F}^{-1}(\mathcal{F}(\mathbf{I}_{f}))\big|$.
\end{itemize}

\paragraph{(3) Stochastic composition (order matters).}
Rather than linearly mixing degradations, we \emph{compose} them in random order to avoid biasing the network to a single corruption sequence:
\begin{equation}
  \mathbf{I}_{d} = \mathcal{T}(\tilde{\mathbf{I}}), \qquad
  \mathcal{T}\in\Big\{
    \underbrace{\mathcal{N}\!\circ\!\mathcal{B}}_{\text{blur $\to$ noise}},
    \underbrace{\mathcal{B}\!\circ\!\mathcal{N}}_{\text{noise $\to$ blur}}
  \Big\},
\end{equation}
where $\mathcal{B}(\cdot)$ is Gaussian blur with $k\in\{3,5,\ldots,17\}$, and
$\mathcal{N}(\cdot)$ is sampled uniformly from the two noise families above.
Concretely, with probability $>0.55$ we apply \emph{blur\,$\to$\,noise} using a random $k$,
and independently with probability $<0.45$ we apply a light \emph{noise\,$\to$\,blur} path
(\emph{Fourier noise} followed by $k{=}3$), so both may occur, yielding a diverse corruption space.
At test time, controlled stress tests are created by setting a user\textendash specified
Fourier noise strength ($\gamma_f\!\in[0,1]$) and/or a fixed blur size $k$.

\paragraph{(4) Targets and normalization.}
All intensities are kept in $[0,1]$ with clipping after each corruption step.
For natural images, the original clean image serves as the target.
For ultrasound scans, clean\textendash like targets are approximated via non\textendash local low\textendash rank denoising:
\begin{equation}
  \mathbf{I}_{t} = \mathcal{D}_{\text{NLLR}}(\mathbf{I}),
\end{equation}
removing the need for ground\textendash truth clean acquisitions.

\medskip
\noindent\textit{Discussion.}
The Gaussian PSF surrogate broadens edges in a controllable manner, while the k\textendash space perturbations inject phase/magnitude fluctuations that manifest as fine\textendash grain texture after inverse FFT, closely resembling the granular statistics observed in ultrasound B\textendash mode. Alternating the operator order (blur\,$\to$\,noise vs.\ noise\,$\to$\,blur) intentionally changes edge\textendash to\textendash noise interactions, preventing the model from overfitting to a single artifact chronology. In practice we find this composition strategy yields superior robustness to unknown scanner settings and consistently improves both fidelity (PSNR) and structure (SSIM) in downstream evaluations.

\subsection{Architecture}
\label{subsec:architecture}
\begin{figure}[!t]
    \centering
    \includegraphics[width=\linewidth]{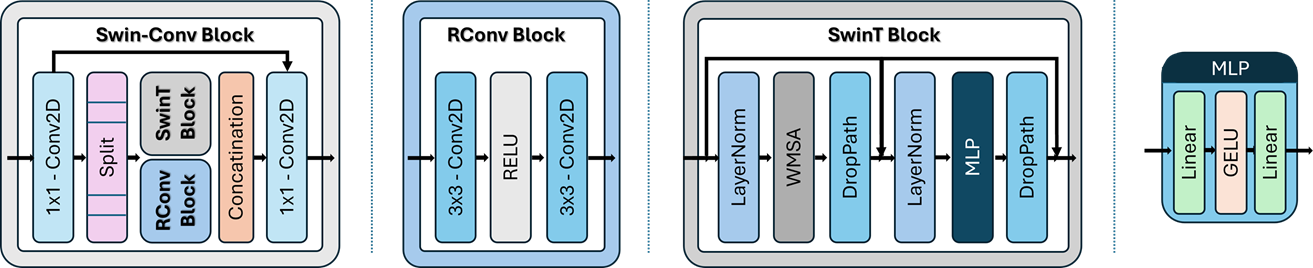}
    \caption{Hybrid Swin-Convolution block. Input channels are split evenly into a local conv path and a shifted–window attention path, fused with $1{\times}1$ projection, and added residually to the input.}
    \label{fig:SwinConvolutionBlock}
\end{figure}
\begin{figure}[!t]
    \centering
    \includegraphics[width=\linewidth]{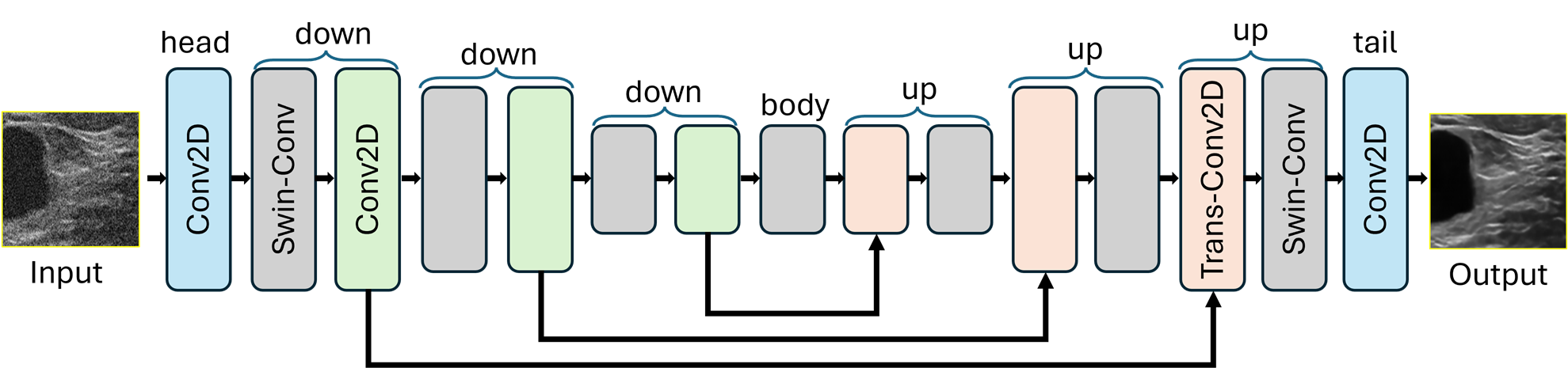}
    \caption{Overall Swin–Conv U–Net. Three encoder stages ($\downarrow$), bottleneck, and three decoder stages ($\uparrow$) with additive cross–scale fusions. Downsampling uses stride–2 conv; upsampling uses stride–2 transposed conv.}
    \label{fig:SCUNET_arch}
\end{figure}

Our backbone is a Swin–Transformer\cite{liu2021swin} augmented U–Net (``Swin–Conv U–Net''), closely following the practical SC-UNet design \cite{zhang2023practical}. It couples local convolutions (robust to speckle statistics) with shifted–window self–attention to capture mid/long range context, and uses additive cross–scale fusions rather than concatenation. Figure~\ref{fig:SwinConvolutionBlock} depicts the core hybrid block; Figure~\ref{fig:SCUNET_arch} shows the overall topology.

\paragraph{Implementation used in this work.}
Unless otherwise stated, we use \texttt{SCUNet(in\_nc{=}1, dim{=}64, config{=}[2,2,2,2,2,2,2])}, window size $M{=}8$, head dimension $d_h{=}32$, and a linear drop–path schedule set to $0$ (disabled). Replication padding brings inputs to the nearest multiple of $64$ on each side; outputs are cropped back to the original size.

\paragraph{Patch embedding and stem.}
A $3{\times}3$ conv maps the grayscale input $\mathbf{I}\!\in\![0,1]^{H\times W\times1}$ to $C{=}64$ channels (``stem'' features). This improves edge fidelity over a purely linear embedding and provides a stable spatial bias for speckle.

\paragraph{Encoder–bottleneck–decoder hierarchy.}
We employ three downsampling stages, a bottleneck, and three symmetric upsampling stages:
\begin{itemize}
  \item \textbf{Down path:} Each stage consists of $N_s$ hybrid blocks (see below) followed by a stride–2 $2{\times}2$ convolution that halves the spatial size and doubles channels: $64\!\to\!128\!\to\!256\!\to\!512$.
  \item \textbf{Bottleneck:} Hybrid blocks operate at the coarsest scale (512 channels) where global context is cheapest.
  \item \textbf{Up path:} Each stage starts with a stride–2 transposed convolution (e.g., $512\!\to\!256$), then hybrid blocks refine features. Instead of U–Net concatenation, we use \emph{additive} fusion with the encoder features at the same scale ($x\!+\!x_{\text{enc}}$); this keeps channel counts fixed and reduces memory without hurting detail.
  \item \textbf{Reconstruction head:} A $3{\times}3$ conv produces the $1$–channel output. We add a final residual with the stem features ($x\!+\!x_{\text{stem}}$) to preserve fine detail and stabilize training.
\end{itemize}

\paragraph{Hybrid Swin–Convolution block (\textit{TransConvBlock}).}
Given an input tensor with $C$ channels at some resolution, we first apply a $1{\times}1$ conv and split channels evenly into a \emph{conv branch} ($C/2$) and a \emph{transformer branch} ($C/2$), process them in parallel, concatenate, project with another $1{\times}1$ conv, and add a residual from the block input:
\[
\mathbf{Y} \;=\; \mathbf{X} \;+\; \underbrace{\phi_{1\times1}\!\Big([\;\underbrace{f_{\text{conv}}(\mathbf{X}_c)}_{\text{local}}\;,\;\underbrace{f_{\text{swin}}(\mathbf{X}_t)}_{\text{nonlocal}}\;]\Big)}_{\text{fusion}}.
\]
The conv branch is a lightweight $3{\times}3$\,–\,ReLU\,–\,$3{\times}3$ residual unit. The transformer branch is a \emph{Swin block} (described below) operating on $(H{\times}W)$ tokens with channel width $C/2$.

\paragraph{Swin block with relative position bias.}
Each block applies LayerNorm and windowed multi–head self–attention (W–MSA) with window size $M=8$ and learnable 2D relative positional bias, followed by an MLP with GELU \cite{hendrycks2016gaussian}; both subpaths are wrapped with residual connections:
\[
\mathbf{Z}=\mathbf{X}+\mathrm{W\text{-}MSA}(\mathrm{LN}(\mathbf{X})),\quad
\mathbf{Y}=\mathbf{Z}+\mathrm{MLP}(\mathrm{LN}(\mathbf{Z})).
\]
We alternate plain and \emph{shifted} windows (\texttt{W}/\texttt{SW}) across consecutive blocks to enable cross–window interaction (\emph{SW} is automatically disabled when the current resolution $\leq M$). The number of heads is determined by the transformer–branch width: $h=\tfrac{C/2}{d_h}$ (e.g., $h=1,2,4,8$ at channel widths $32,64,128,256$). 


\subsection{Datasets}
\begin{figure}[!htbp]
    \centering
    \includegraphics[width=\linewidth]{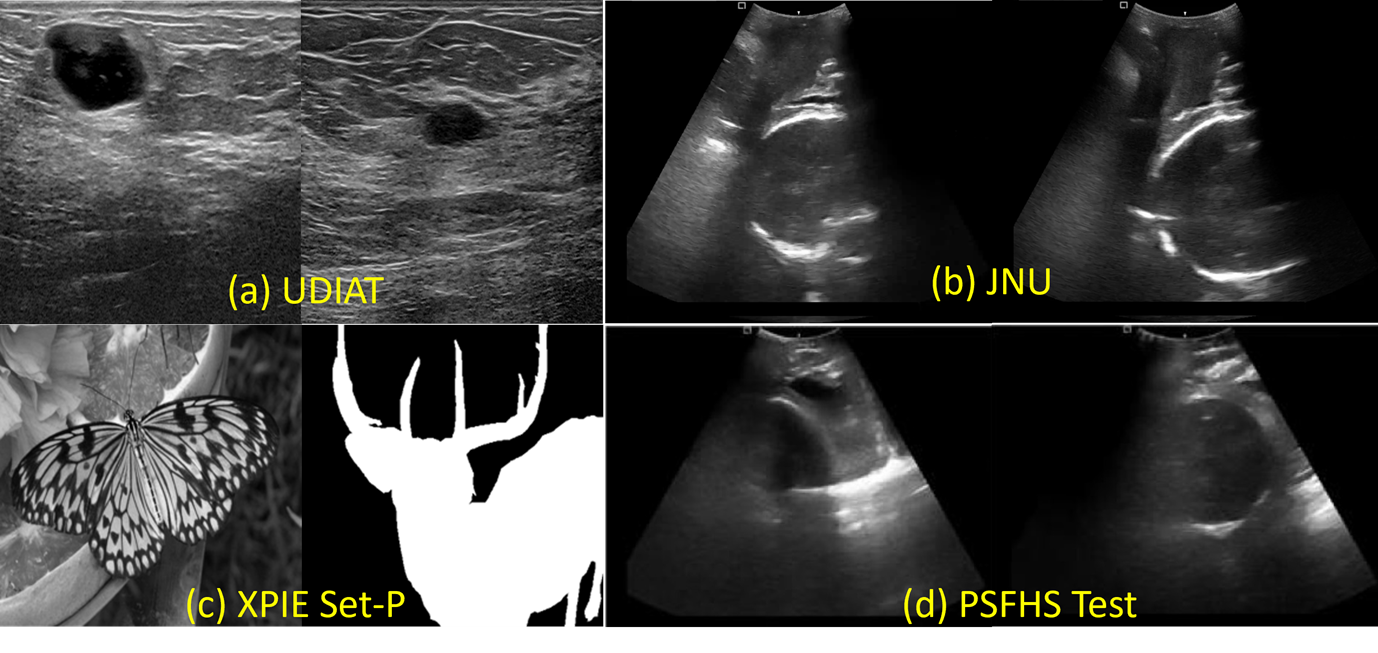}
    \caption{Sample images from datasets: (a) UDIAT, (b) JNU-IFM, (c) XPIE Set-P, and (d) PSFHS Test.}
    \label{fig:dataset_samples}
\end{figure}
We evaluate the proposed method using three public datasets for training/validation and one benchmark dataset for downstream segmentation. Unless noted, all images are used in grayscale B-mode.

\begin{itemize}
    \item \textbf{UDIAT Dataset B} \cite{yap2018udaitb}: Breast ultrasound dataset with tumor annotations, collected at the UDIAT Diagnostic Center, Spain, using a Siemens ACUSON scanner. The average image size is $760 \times 570$ pixels. Publicly available at \url{https://helward.mmu.ac.uk/STAFF/m.yap/dataset.php}. Samples used: 131 (train), 16 (val), 16 (test).

    \item \textbf{JNU-IFM} \cite{lu2022jnu}: Intrapartum fetal monitoring dataset with grayscale 2D transperineal ultrasound scans. It contains 6224 images extracted from 78 videos of 51 patients, acquired using a Youkey D8 wireless probe. Labels were validated by expert radiologists. Available at \url{https://figshare.com/articles/dataset/JNU-IFM/14371652}.  Samples used: 4224 (train), 1000 (val), 1000 (test).

    \item \textbf{XPIE Set-P} \cite{xia2017xpie}: Natural images with ground-truth masks used to expose the model to cross-domain textures and to synthesize phantom-like inputs during degradation modeling. Available at \url{http://cvteam.net/projects/CVPR17-ELE/XPIE.tar.gz}. \textit{Split:} 850 (train), 200 (val), 200 (test).

    \item \textbf{PSFHS} \cite{chen2024psfhs}: Intrapartum transperineal ultrasound annotated for pubic symphysis and fetal head; image size $256\times256$ pixels. Access upon request via \url{https://ps-fh-aop-2023.grand-challenge.org/}. \textit{Split used in this work:} 3200 (train), 800 (val), 700 (test). \textbf{Important:} The PSFHS \emph{train/val} splits are used \emph{only} to train the downstream UNet segmentation model; the proposed denoising model is \emph{not} trained on PSFHS. The 700-image test split is held out exclusively for reporting segmentation with/without our preprocessing.    
\end{itemize}

Sample images from each dataset is shown in Figure~\ref{fig:dataset_samples}.

\subsection{Quantitative Metrics}\label{sec:quantitative_metrics}
We report both \emph{resolution-recovery} and \emph{perceptual/fidelity} metrics. Resolution metrics are computed from 1D intensity profiles extracted from regions of interest (ROIs), while quality metrics are evaluated on full images. \footnote{Rule-of-thumb bands below are indicative and depend on sampling, anatomy, and acquisition. When in doubt, compare \emph{relative} improvements to the input or a non-degraded reference.}

\subsubsection{Resolution Metrics}
\begin{itemize}
\item \textbf{Full Width at Half Maximum (FWHM):}
For a 1D profile $p(x)$, the FWHM is the width at half the maximum:
\begin{equation}
\mathrm{FWHM} = x_2 - x_1,\quad p(x_1)=p(x_2)=\tfrac{1}{2}\max_x p(x).
\end{equation}
Lower values indicate sharper transitions.

\item \textbf{Mean Gradient (GradMean):}
\begin{equation}
    \mathrm{GradMean} = \frac{1}{N}\sum_{i=1}^{N}\bigl|\nabla p(x_i)\bigr|.
\end{equation}
Higher is sharper on average.
\emph{Rule-of-thumb (images scaled to $[0,1]$):} \textbf{ideal} $\ge 0.05$, \textbf{good} $0.03\!-\!0.05$, \textbf{poor} $\le 0.02$.

\item \textbf{Maximum Gradient (GradMax):}
\begin{equation}
    \mathrm{GradMax} = \max_{i}\bigl|\nabla p(x_i)\bigr|.
\end{equation}
Captures the steepest edge.
\emph{Rule-of-thumb ($[0,1]$ scale):} \textbf{ideal} $\ge 0.30$, \textbf{good} $0.18\!-\!0.30$, \textbf{poor} $\le 0.12$.

\item \textbf{Contrast:}
\begin{equation}
    \mathrm{Contrast} = \frac{I_{\max}-I_{\min}}{I_{\max}+I_{\min}},
\end{equation}
with $I_{\max}, I_{\min}$ the extremal intensities in the ROI; higher is better.
\emph{Rule-of-thumb:} \textbf{ideal} $\ge 0.95$, \textbf{good} $0.90\!-\!0.95$, \textbf{poor} $\le 0.80$.
\end{itemize}

\subsubsection{Quality Metrics}
\begin{itemize}
\item \textbf{Peak Signal-to-Noise Ratio (PSNR):}
\begin{equation}
\mathrm{PSNR} = 10\log_{10}\left(\frac{L^2}{\mathrm{MSE}}\right),
\end{equation}
where $L$ is the dynamic range and $\mathrm{MSE}$ is the mean squared error between the reference and the reconstruction.
\textbf{In all our experiments we use $L{=}255$} (8-bit range).%
\footnote{If images are normalized to $[0,1]$, then $L{=}1$. We explicitly rescale to the 8-bit range for consistent reporting.}

\item \textbf{Structural Similarity Index (SSIM):}
For patches $x$ and $y$,
\begin{equation}
    \mathrm{SSIM}(x,y)=\frac{(2\mu_x\mu_y + C_1)(2\sigma_{xy} + C_2)}{(\mu_x^2+\mu_y^2 + C_1)(\sigma_x^2+\sigma_y^2 + C_2)},
\end{equation}
where $\mu_x, \mu_y$ are means, $\sigma_x^2, \sigma_y^2$ variances, and $\sigma_{xy}$ the covariance.
In the standard Python implementations, the stabilizers are set via
\[
    C_1=(k_1 L)^2,\qquad C_2=(k_2 L)^2,
\]
with \textbf{$k_1{=}0.01$, $k_2{=}0.03$} and the same \textbf{$L{=}255$} used for PSNR.
\end{itemize}

\subsection{Training and Evaluation Protocol}
We train the network to map a degraded input $\mathbf{I}_{d}$ to an enhanced image $\hat{\mathbf{I}} = f_{\theta}(\mathbf{I}_{d})$ using paired targets $\mathbf{I}_{t}$ (natural-image originals or NLLR-filtered ultrasound). The objective is a pixelwise $\ell_{1}$ loss:
\begin{equation}
    \mathcal{L}(\theta) \;=\; \bigl\|\, f_{\theta}(\mathbf{I}_{d}) - \mathbf{I}_{t} \,\bigr\|_{1}.
\end{equation}

\noindent\textbf{Hyperparameters and implementation.}
We use \texttt{num\_epochs}$=4000$, \texttt{learning\_rate}$=10^{-4}$, and \texttt{batch\_size}$=16$. Mini-batches are shuffled every epoch. To accommodate the encoder–decoder strides, inputs are padded along height and width to the nearest multiple of $64$ and cropped back after inference (no tiling). Degraded inputs $\mathbf{I}_{d}$ are generated on-the-fly via the physics-guided pipeline (random PSF blur, additive Gaussian noise, and Fourier-domain perturbations), etc.

\noindent\textbf{Training data and targets.}
The training set contains \textbf{5205} images in total: UDIAT~B (\textbf{131} breast ultrasound images), JNU-IFM (\textbf{4224} intrapartum transperineal ultrasound frames), and \textbf{850} natural images (with masks) from XPIE Set-P used to synthesize phantom-like textures. For natural images, the clean original serves as $\mathbf{I}_t$; for ultrasound images, clean-like targets are obtained using non-local low-rank denoising~\cite{zhu2017non}, i.e., $\mathbf{I}_{t} = \mathcal{D}_{\text{NLLR}}(\mathbf{I})$.

\noindent\textbf{Model selection and validation.}
Because ground-truth clean B-mode is unavailable, checkpoint selection combines (i) PSNR/SSIM on validation pairs (natural images and NLLR targets) with (ii) visual inspection on ultrasound validation images (edge sharpness, speckle suppression, artifact avoidance). This composite criterion emphasizes clinical plausibility over any single surrogate metric.

\noindent\textbf{Segmentation-based evaluation.}
To quantify downstream utility, we train a UNet\cite{ronneberger2015unet} on the PSFHS \emph{training/validation} split (fetal head and pubic symphysis) and evaluate it \emph{with vs.\ without} the proposed preprocessing. PSFHS images are \emph{not} used to train the denoiser. We report Dice on the independent \textbf{700}-image PSFHS \emph{test} set alongside the image-quality metrics from Section~\ref{sec:quantitative_metrics}, thereby assessing domain generalization and the impact on anatomical delineation.

\section{RESULTS AND DISCUSSION}
This section evaluates proposed method qualitatively and quantitatively. We begin with a controlled resolution-recovery study (Fig.~\ref{fig:speckle_resolution_comparison}) under increasing PSF blur (Sec.~\ref{sec:speckle_resolution_eval}), then present cross-dataset deblurring results (Fig.~\ref{fig:blur_samples}, Table~\ref{tab:blur_psnr_ssim_merged}). Next, we analyze generalized denoising under Gaussian and speckle noise ladders (Figs.~\ref{fig:new_gaussian_psnr}--\ref{fig:new_speckle_ssim}), followed by qualitative noise-to-denoise visualization (Fig.~\ref{fig:noise_denoise_grid}). Finally, we report results on combined blur and noise modeling and downstream segmentation robustness (Figs.~\ref{fig:noisy_blur}, \ref{fig:dice_score_vs_noise}). 

\subsection{Evaluation of Spatial Resolution Recovery}
\label{sec:speckle_resolution_eval}
\begin{figure*}[ht]
    \centering
    \includegraphics[width=0.9\linewidth]{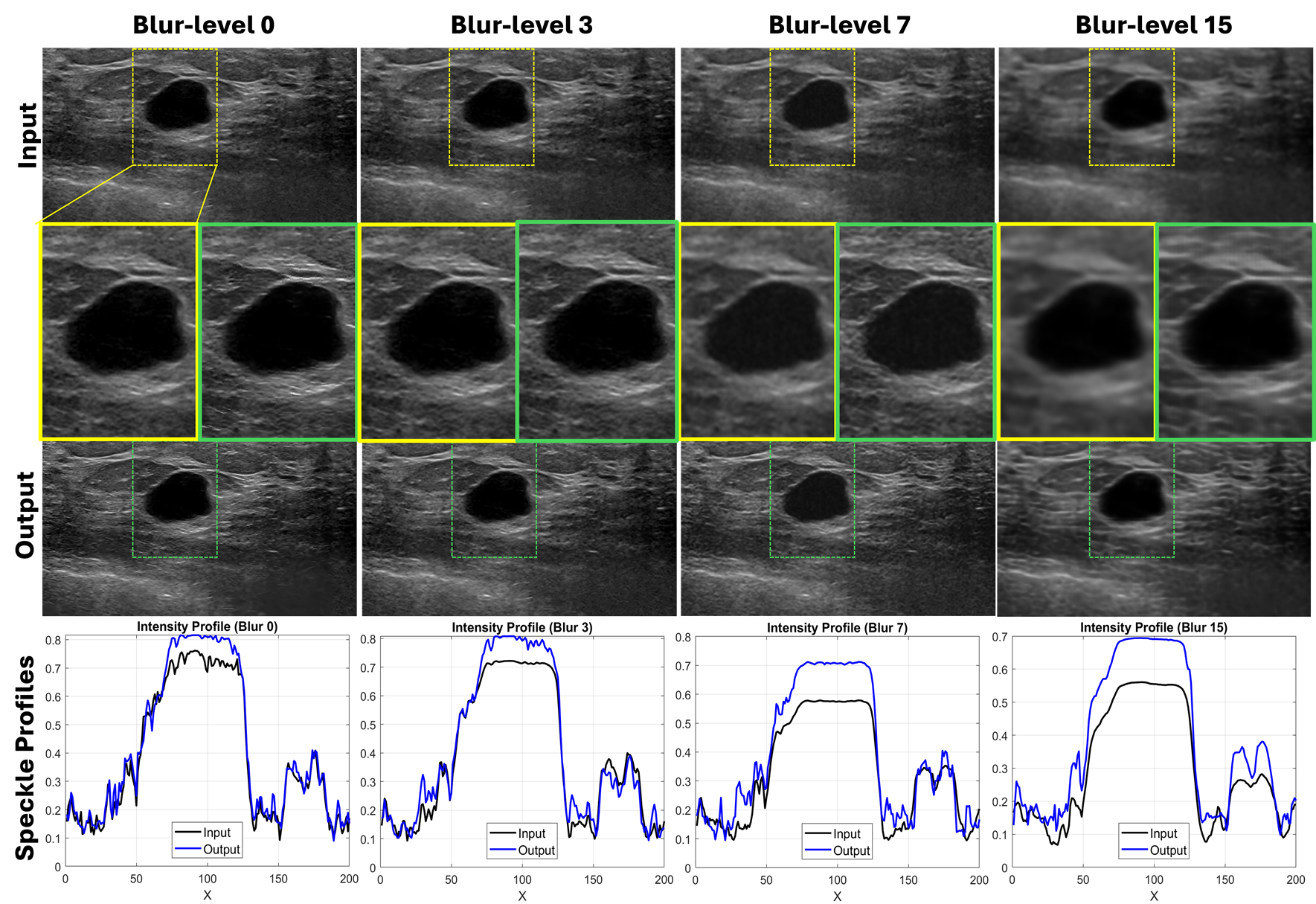}
    \caption{Resolution recovery under increasing PSF blur ($\sigma=0,3,7,15$). First/Third row: full frames; second row: ROI zoom-ins; fourth row: 1D intensity profiles comparing inputs vs.\ proposed method outputs.}
    \label{fig:speckle_resolution_comparison}
\end{figure*}

We conducted a controlled experiment to quantify spatial resolution recovery at four blur levels $\sigma \in \{0,3,7,15\}$ (Gaussian PSF). For each case, we extracted a vertical ROI ($x=1\!:\!200$, $y=195\!:\!200$), computed mean intensity profiles, and measured FWHM, GradMean, GradMax, and Contrast (Section~\ref{sec:quantitative_metrics}). Qualitative comparisons and profile plots are shown in Fig.~\ref{fig:speckle_resolution_comparison}. Figure~\ref{fig:speckle_resolution_comparison} shows, per blur level, the full frame (first and third row), ROI zoom (second row), and input vs.\ proposed profile overlays (fourth row), revealing progressive edge sharpening as blur increases.

\begin{table}[!htbp]
\centering
\caption{Resolution and sharpness metrics across blur levels. ``Yes'' indicates proposed method reconstruction.}
\label{tab:resolution_metrics}
\resizebox{\linewidth}{!}{
\begin{tabular}{|c|c|c|c|c|c|}
\hline
\textbf{Blur $\sigma$} & \textbf{Recon} & \textbf{FWHM} & \textbf{GradMean} & \textbf{GradMax} & \textbf{Contrast} \\
\hline
0  & No  & 141 & 0.0533 & 0.2529 & 0.9087 \\
0  & Yes & 138 & 0.0558 & 0.3536 & 0.9500 \\
\hline
3  & No  & 142 & 0.0410 & 0.1850 & 0.8948 \\
3  & Yes & 138 & 0.0443 & 0.2974 & 0.9149 \\
\hline
7  & No  & 144 & 0.0246 & 0.1641 & 0.8824 \\
7  & Yes & 141 & 0.0397 & 0.2386 & 0.9009 \\
\hline
15 & No  & 145 & 0.0146 & 0.1007 & 0.8675 \\
15 & Yes & 142 & 0.0324 & 0.1752 & 0.8941 \\
\hline
\end{tabular}}
\end{table}

\noindent\textbf{Apparent super-resolution.} Notably, even at $\sigma\!=\!0$ the output profile exhibits a slightly narrower FWHM and higher peak gradient than the input, suggesting a mild deblurring of scanner-limited blur and an \emph{apparent} super-resolution effect in the 1D profile sense, achieved without hallucinating structures. For moderate blur ($\sigma\!=\!3,7$), GradMax increases notably (e.g., $0.2974$ vs.\ $0.1850$ at $\sigma\!=\!3$), and FWHM approaches the unblurred case, evidencing edge restoration. Under severe blur ($\sigma\!=\!15$), FWHM and gradient statistics recover substantially, with a modest contrast trade-off that remains visually acceptable.

\subsection{Cross-Dataset Deblurring: Visuals and PSNR/SSIM}
\begin{figure}[!ht]
    \centering
    \includegraphics[width=0.9\linewidth]{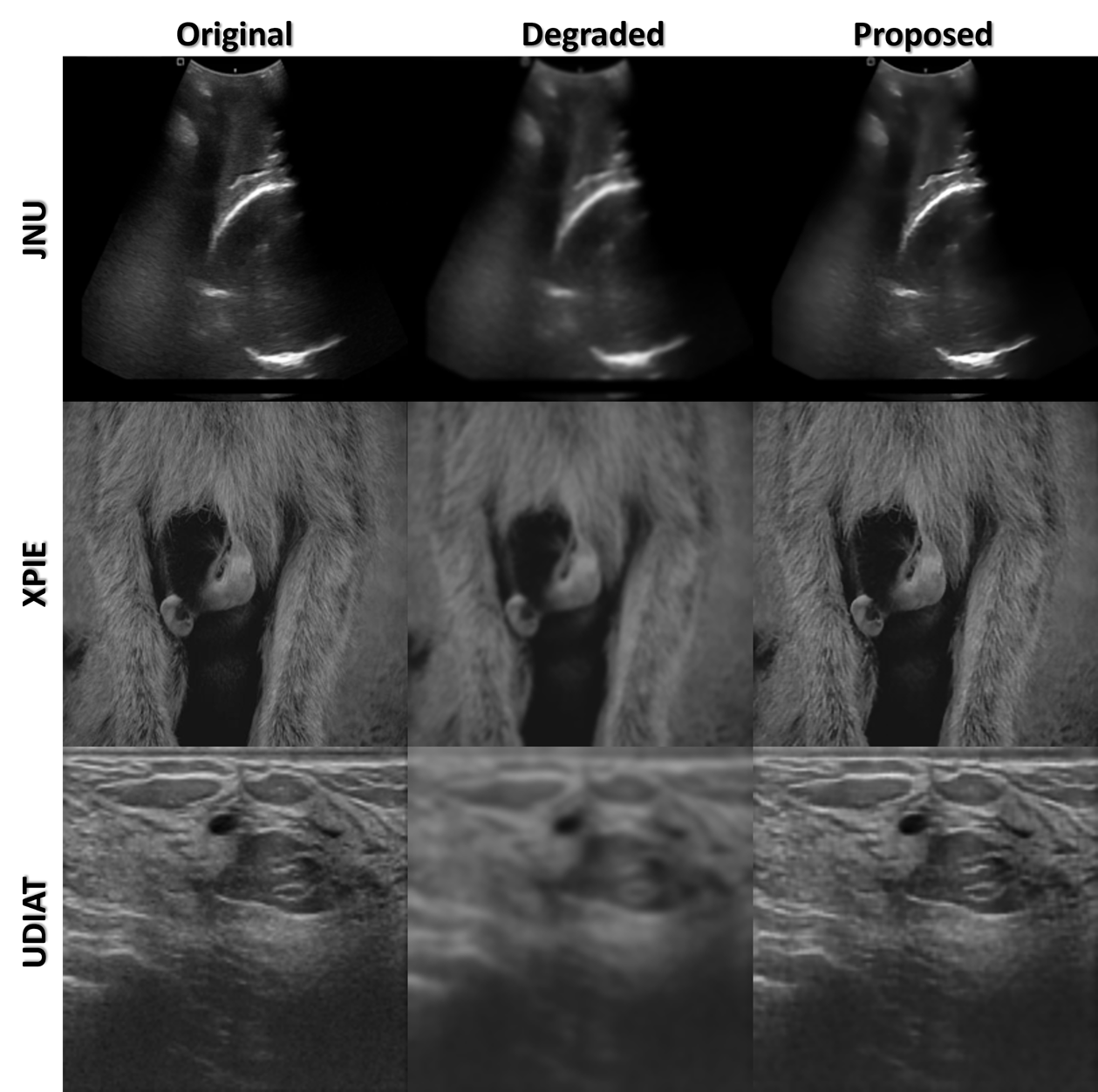}
    \caption{Single-level comparison (blur level $\sigma{=}7$) across datasets showing original, blurred, and proposed deblurred outputs.}
    \label{fig:blur_samples}
\end{figure}
\begin{table*}[!htbp]
\centering
\caption{Effect of PSF blur level ($\sigma$) on image quality across datasets. Each cell shows \textbf{PSNR/SSIM} with two-decimal precision (PSNR in dB). $\infty$ indicates zero MSE.}
\label{tab:blur_psnr_ssim_merged}
\resizebox{\linewidth}{!}{
\setlength{\tabcolsep}{4pt}
\begin{tabular}{|c|cc|cc|cc|cc|cc|}
\hline
\multirow{2}{*}{\textbf{Blur $\sigma$}} & \multicolumn{2}{c|}{\textbf{JNU}} & \multicolumn{2}{c|}{\textbf{UDIAT}} & \multicolumn{2}{c|}{\textbf{XPIE Object}} & \multicolumn{2}{c|}{\textbf{XPIE Mask}} & \multicolumn{2}{c|}{\textbf{PSFHS Test}} \\
\cline{2-11}
& \textbf{Input} & \textbf{Output} & \textbf{Input} & \textbf{Output} & \textbf{Input} & \textbf{Output} & \textbf{Input} & \textbf{Output} & \textbf{Input} & \textbf{Output} \\
\hline
0  & $\infty$/1.00 & $\infty$/1.00 & $\infty$/1.00 & $\infty$/1.00 & $\infty$/1.00 & $\infty$/1.00 & $\infty$/1.00 & $\infty$/1.00 & $\infty$/1.00 & $\infty$/1.00 \\
3  & 40.05/0.98 & 53.10/1.00 & 30.82/0.90 & 40.73/0.99 & 33.11/0.94 & 40.91/0.99 & 33.14/0.99 & 44.90/1.00 & 32.25/0.93 & 39.99/0.98 \\
5  & 37.10/0.96 & 45.86/1.00 & 28.24/0.82 & 35.26/0.96 & 30.30/0.88 & 37.02/0.97 & 30.39/0.98 & 41.86/1.00 & 30.19/0.89 & 35.08/0.96 \\
7  & 34.79/0.94 & 43.41/0.99 & 26.42/0.73 & 34.92/0.97 & 28.13/0.82 & 36.43/0.97 & 28.11/0.97 & 40.85/1.00 & 28.54/0.85 & 35.75/0.97 \\
9  & 33.65/0.93 & 43.72/0.99 & 25.63/0.68 & 32.57/0.94 & 27.06/0.77 & 34.97/0.95 & 26.92/0.95 & 40.22/1.00 & 27.70/0.82 & 34.63/0.96 \\
11 & 32.71/0.91 & 41.80/0.99 & 25.03/0.63 & 29.85/0.89 & 26.20/0.74 & 32.59/0.92 & 25.91/0.94 & 39.46/1.00 & 26.99/0.80 & 33.18/0.95 \\
13 & 31.94/0.90 & 40.58/0.99 & 24.57/0.60 & 28.64/0.86 & 25.51/0.70 & 31.27/0.90 & 25.07/0.93 & 38.84/1.00 & 26.39/0.78 & 31.88/0.93 \\
15 & 31.30/0.89 & 39.59/0.99 & 24.21/0.57 & 27.77/0.83 & 24.94/0.68 & 30.29/0.87 & 24.37/0.92 & 38.15/1.00 & 25.88/0.76 & 30.81/0.92 \\
\hline
\end{tabular}
}
\end{table*}

Figure~\ref{fig:blur_samples} shows single-level ($\sigma{=}7$) examples across JNU, UDIAT, and XPIE, where proposed method visibly restores edges and suppresses speckle-like texture. Table~\ref{tab:blur_psnr_ssim_merged} summarizes PSNR/SSIM vs.\ blur level for inputs and proposed outputs across datasets. Gains are consistent from mild to severe blur, indicating robustness to PSF variation.

\noindent\textbf{Quantitative gains.}
Across the blur ladder ($\sigma\!\in\!\{3,\dots,15\}$) and all five datasets, the proposed method raises PSNR by \textbf{+3.56} to \textbf{+13.78}\,dB and SSIM by \textbf{+0.01} to \textbf{+0.26} (Table~\ref{tab:blur_psnr_ssim_merged}).
At the operating point used in Fig.~\ref{fig:blur_samples} ($\sigma{=}7$), the mean improvement over all datasets is \textbf{+9.07}\,dB PSNR and \textbf{+0.118} SSIM, with per–dataset gains:
JNU \textit{(+8.62\,dB / +0.05)}, UDIAT \textit{(+8.50\,dB / +0.24)}, XPIE–Object \textit{(+8.30\,dB / +0.15)}, XPIE–Mask \textit{(+12.74\,dB / +0.03)}, and PSFHS \textit{(+7.21\,dB / +0.12)}.
The largest PSNR uplift occurs on XPIE–Mask at severe blur ($\sigma{=}15$: \textbf{+13.78}\,dB), while the largest SSIM gains are on UDIAT (up to \textbf{+0.26} for $\sigma\!\ge\!9$).
Overall, improvements persist from mild to heavy blur, and they \emph{increase relatively} as blur severity grows, indicating robustness to PSF variation and consistent recovery of both fidelity (PSNR) and perceptual structure (SSIM).

\subsection{Generalized Denoising: Quantitative and Qualitative Analyses}
\begin{figure}[!htbp]
    \centering
    \includegraphics[width=\linewidth]{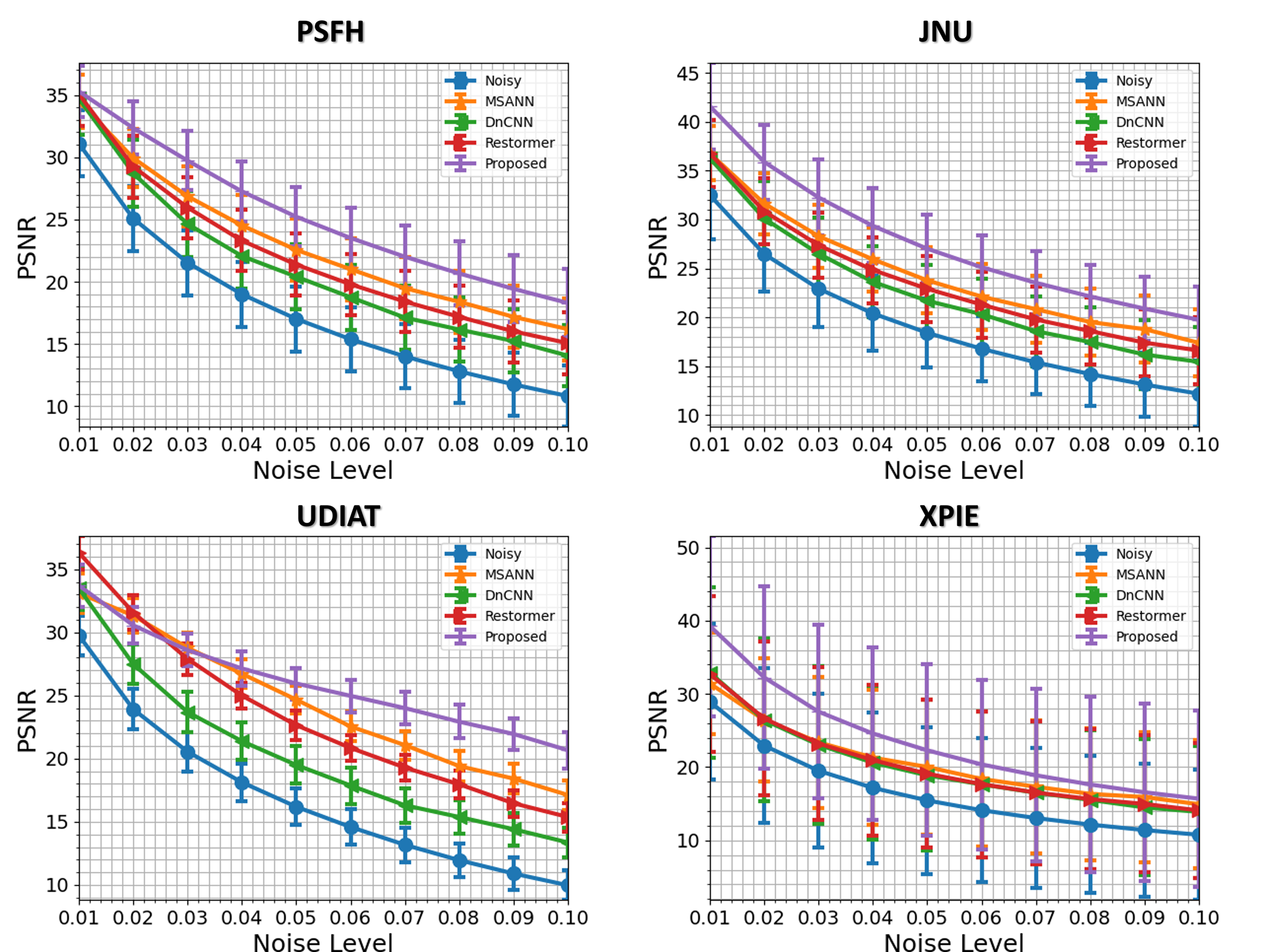}
    \caption{PSNR vs.\ Gaussian noise level across datasets.}
    \label{fig:new_gaussian_psnr}
\end{figure}

\begin{figure}[!htbp]
    \centering
    \includegraphics[width=\linewidth]{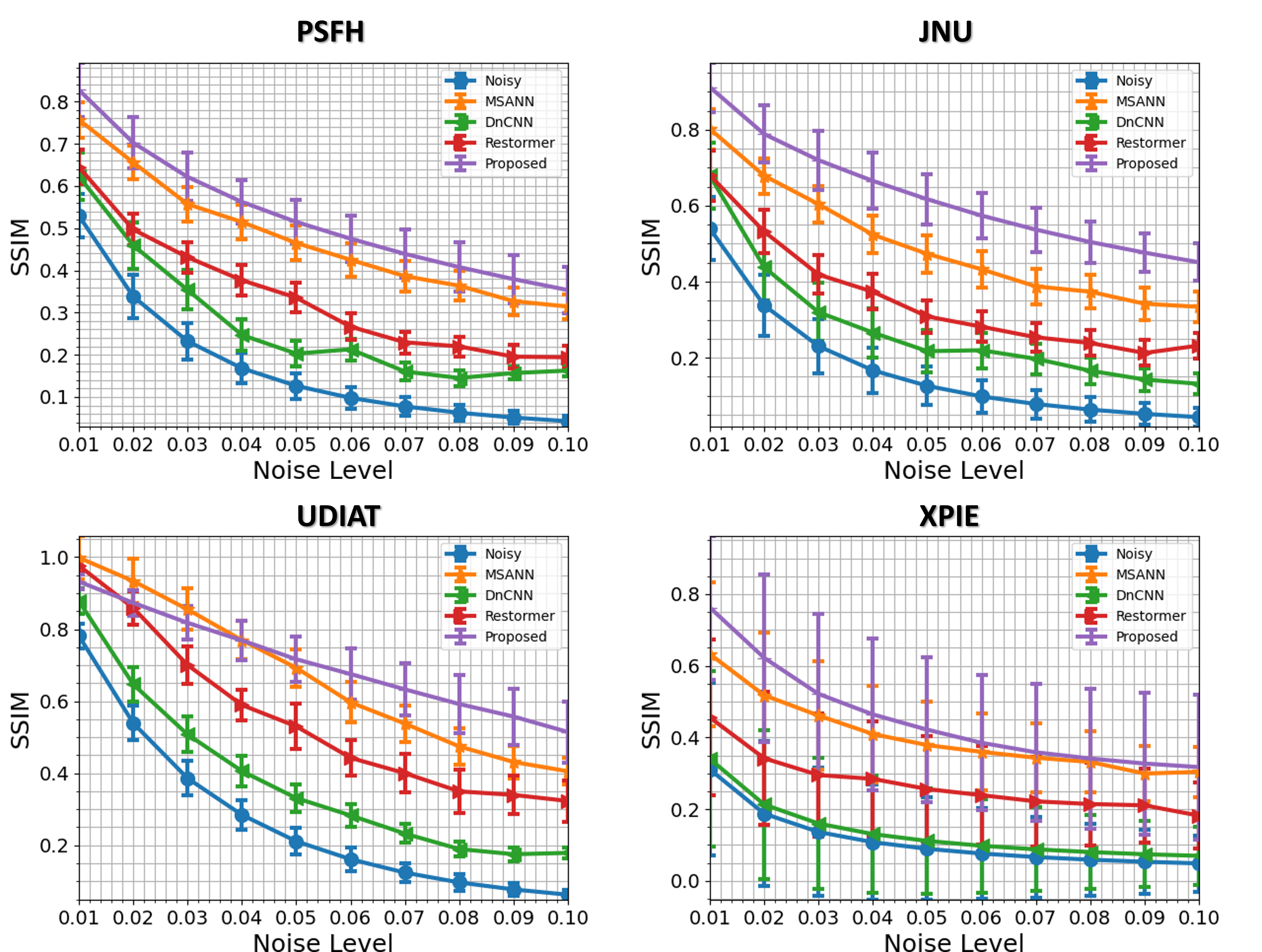}
    \caption{SSIM vs.\ Gaussian noise level across datasets.}
    \label{fig:new_gaussian_ssim}
\end{figure}

\begin{figure}[!htbp]
    \centering
    \includegraphics[width=\linewidth]{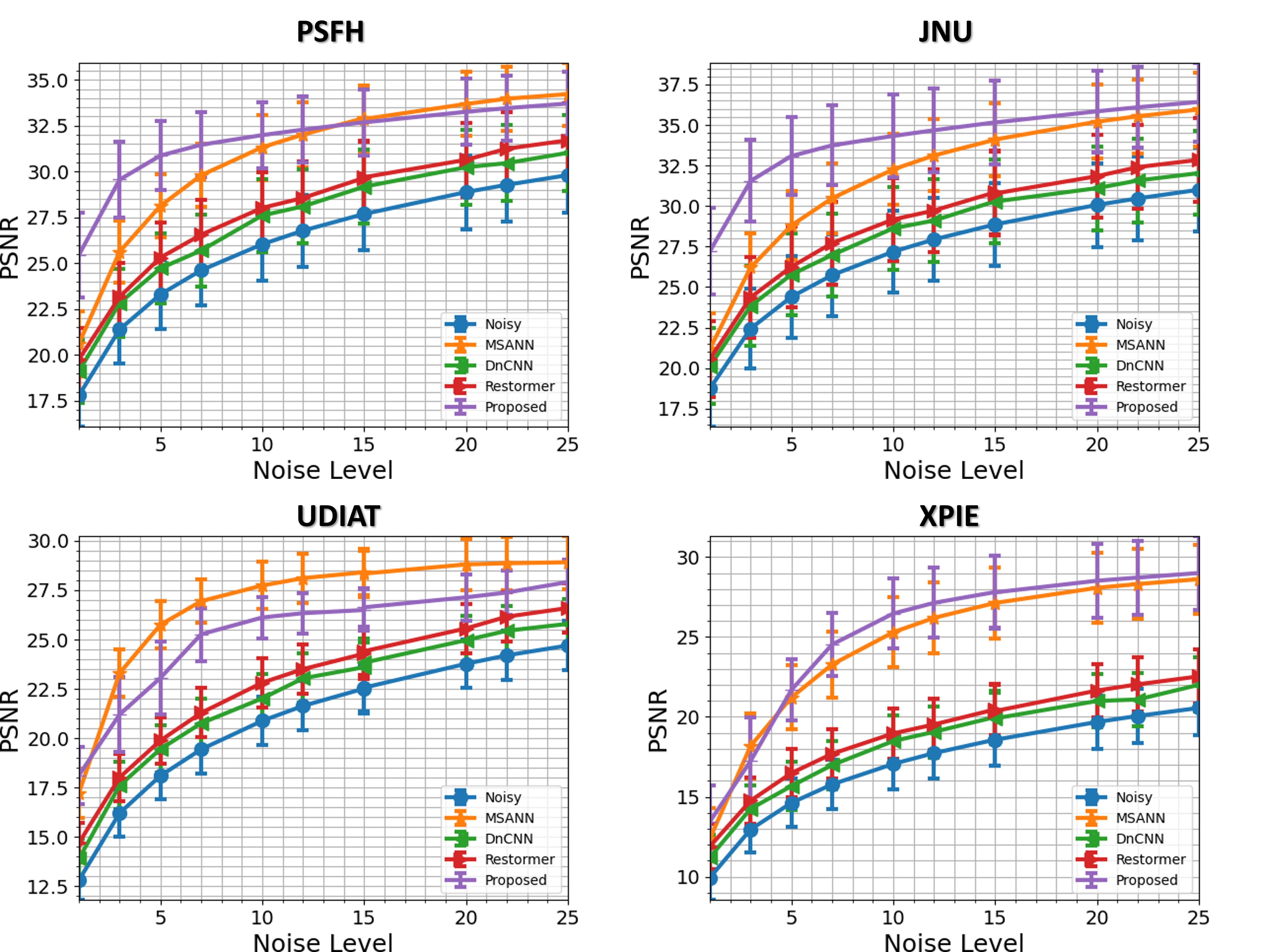}
    \caption{PSNR vs.\ speckle noise level across datasets.}
    \label{fig:new_speckle_psnr}
\end{figure}

\begin{figure}[!htbp]
    \centering
    \includegraphics[width=\linewidth]{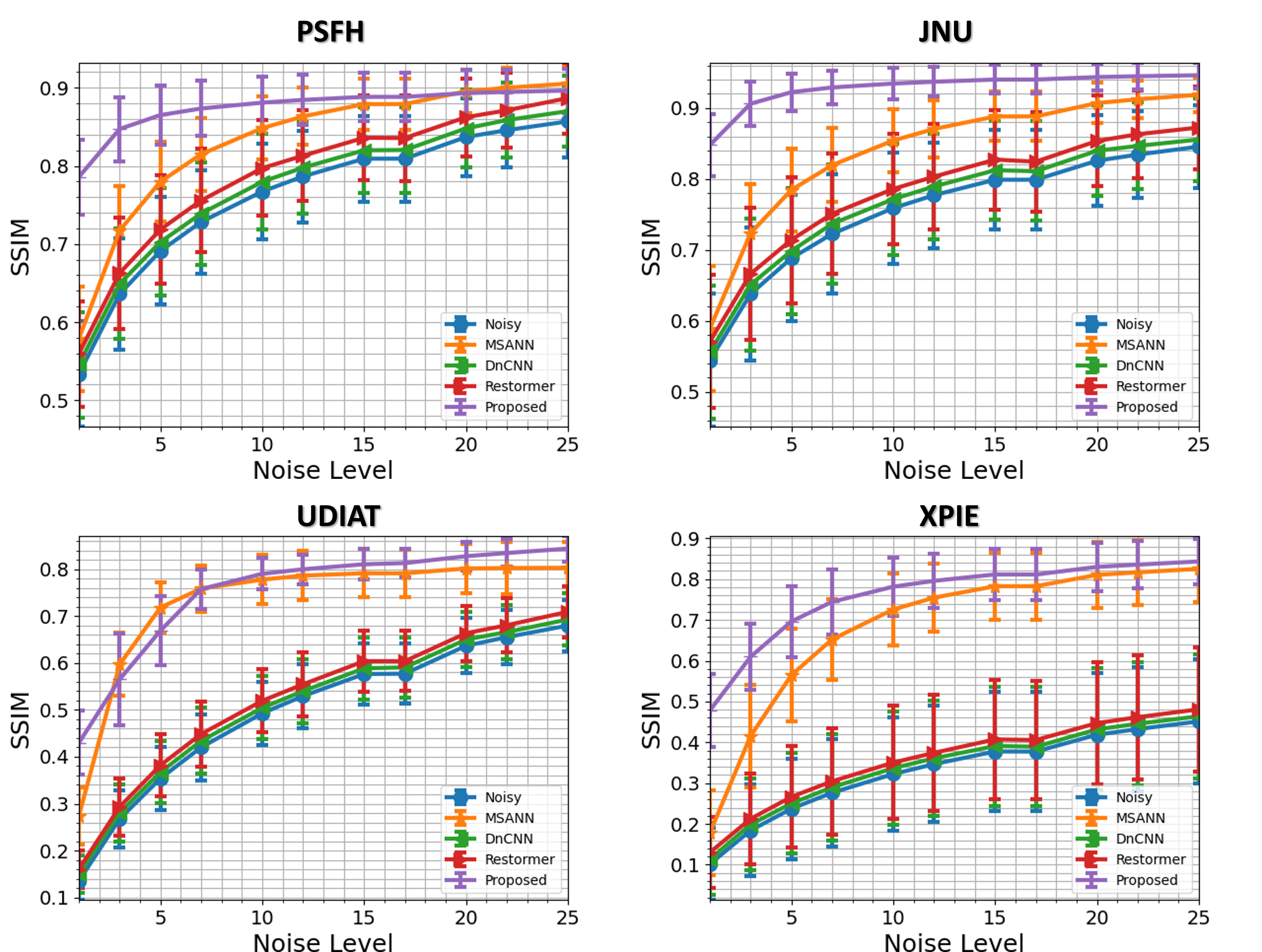}
    \caption{SSIM vs.\ speckle noise level across datasets.}
    \label{fig:new_speckle_ssim}
\end{figure}

\noindent\textbf{Noise ladders (Gaussian \& speckle).}
Figures~\ref{fig:new_gaussian_psnr}--\ref{fig:new_speckle_ssim} compare proposed method with MSANN~\cite{guo2023blind}, Restormer~\cite{zamir2022restormer}, and DnCNN~\cite{zhang2017beyond} across five datasets under 11 Gaussian levels ($\sigma\!\in[0.01,0.10]$) and 11 speckle levels $\mathbf{L} \in\{1,3,5,7,10,12,15,17,20,22,25\}$.
For a fair comparison the speckle noise model for test scenario is matched with MSANN~\cite{guo2023blind}, Restormer~\cite{zamir2022restormer}, and DnCNN~\cite{zhang2017beyond}. 
 The multiplicative speckle noise model for test is defined as follows:
    \begin{equation}
        \mathbf{I}_{n} = \mathbf{I} \odot \mathbf{n},
    \end{equation}  
    where $\mathbf{I}_{n}$ represents the noisy image and $\odot$ denotes element-wise multiplication and $\mathbf{n}$ follows  Gamma distribution~\cite{guo2023blind}. The corresponding probability density function $p(\mathbf{n})$ is equivalent to the following formula:
    \begin{equation}
        p(\mathbf{n}) = \frac{\mathbf{L}^\mathbf{L}\mathbf{n}^{(\mathbf{L}-1)}e^{-\mathbf{L}\mathbf{n}}}{\Gamma(\mathbf{L})},
    \end{equation}
    where $\mathbf{L}$ is the noise level, which is also named as equivalent number of looks (ENL), and $\Gamma(.)$ denotes the Gamma function with $\mathbf{n} \ge 0$, $\mathbf{L} \ge 1$.


\emph{Gaussian noise (Figs.~\ref{fig:new_gaussian_psnr}--\ref{fig:new_gaussian_ssim}).}
As expected, PSNR and SSIM decay monotonically with increasing $\sigma$.
Across all datasets, proposed method stays on top for almost the entire ladder and the gap widens in the mid/high-noise regime ($\sigma\!\ge\!0.06$).
At these harder settings, proposed method typically preserves an additional $\sim$1–4\,dB PSNR over Restormer and $\sim$2–6\,dB over DnCNN/MSANN, with SSIM gains of roughly 0.05–0.15.
Interestingly, even on the cross-domain XPIE Object/Mask sets proposed method shows similar performance gain, indicating stronger generalization.
Error bars are consistently small for proposed method, especially at higher noise, suggesting more stable behavior across images.

\emph{Speckle noise (Figs.~\ref{fig:new_speckle_psnr}--\ref{fig:new_speckle_ssim}).}
Here, higher values of $\mathbf{L}$ imply milder speckle, so PSNR/SSIM rise and then saturate. The proposed method shows the steepest early gains in the hardest regime (small $\mathbf{L}$) and reaches a higher plateau than competing methods. \emph{Except on UDIAT}, where MSANN attains the best PSNR, the typical margin in the severe-noise regime is on the order of $\sim$2--5\,dB PSNR and 0.05--0.20 SSIM over the next-best model, with the gap persisting though narrowing as $\mathbf{L}$ increases. The consistent performance gains across datasets underscore robustness beyond the majority distribution. Overall, the ladders show that the proposed method does not merely denoise “on average”; it consistently shifts the entire PSNR/SSIM curve upward, with the largest upward shift where the problem is hardest.

\begin{figure}[!htbp]
    \centering
    \includegraphics[width=1\linewidth]{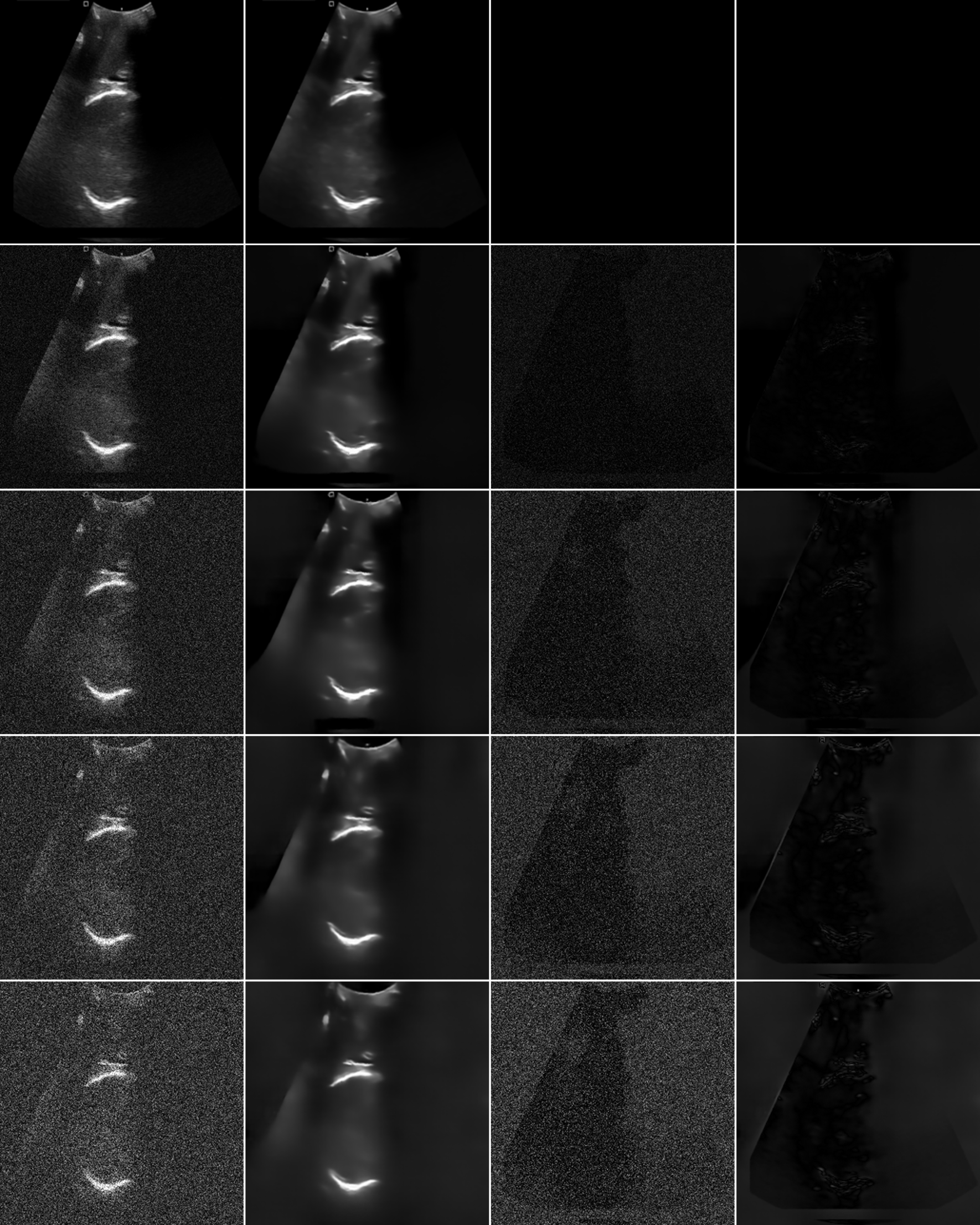}
    \caption{\textbf{Noise-to-denoise grid across Gaussian noise levels.}
    Rows (top$\to$bottom): $\sigma\in\{0,\,0.03,\,0.06,\,0.08,\,0.10\}$.
    Columns (left$\to$right): degraded input, proposed method output (proposed), input residual, output residual.
    Residuals are absolute differences to the reference (non-degraded) image, shown with a shared dynamic range across rows.}
    \label{fig:noise_denoise_grid}
\end{figure}
\noindent\textbf{Visual noise suppression.}

Figure~\ref{fig:noise_denoise_grid} shows that as $\sigma$ increases from $0$ to $0.10$, the input residuals become dominated by fine-grained speckle across the field of view, whereas proposed method outputs retain clean, continuous boundaries. The output residuals concentrate around true edges with markedly lower background energy, indicating effective noise removal without edge erosion or haloing. This qualitative trend is consistent across all rows and aligns with the quantitative gains reported on the Gaussian ladders.

\subsection{Combined Blur and Noise and Downstream Segmentation}
\begin{figure*}[!ht]
    \centering
    \includegraphics[width=0.9\textwidth]{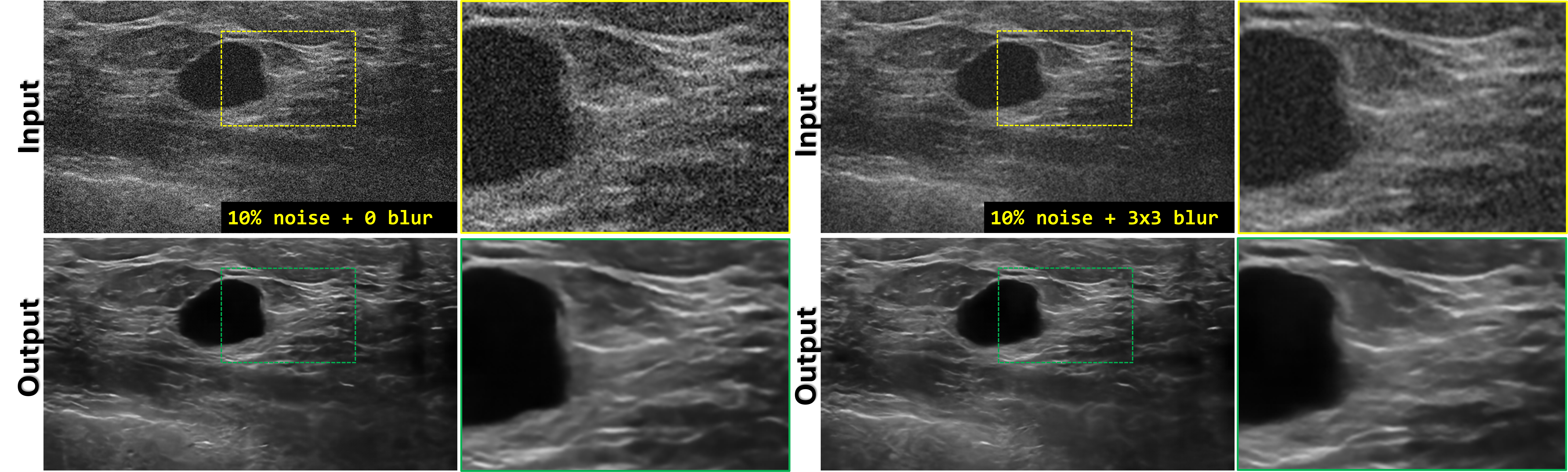}
    \caption{Inference-time comparison under compound degradations (same trained model, no tuning). 
    Left: noisy inputs without additional PSF-like blur; right: blur{+}noise inputs. 
    Rows show inputs, outputs, and zoomed ROIs. The proposed method yields low residual speckle and sharper edges in both conditions.}
    \label{fig:noisy_blur}
\end{figure*}
\noindent\textbf{Joint deblurring and denoising at inference.}
We assess robustness to compound degradations by evaluating the \emph{trained} model (no test-time tuning) on held-out images under two inference conditions: (i) noisy inputs without additional PSF-like blur and (ii) inputs with simultaneous noise and PSF-like blur. As shown in Fig.~\ref{fig:noisy_blur}, the proposed method consistently suppresses background speckle and restores crisp boundaries in both cases. These qualitative trends align with the quantitative resolution statistics in Table~\ref{tab:resolution_metrics}—reduced FWHM with higher GradMax/GradMean and competitive contrast—without haloing or over-sharpening. The same model is used across all panels, and no retraining or hyperparameter changes are applied between conditions.

\begin{figure}[!ht]
    \centering
    \includegraphics[width=0.9\linewidth]{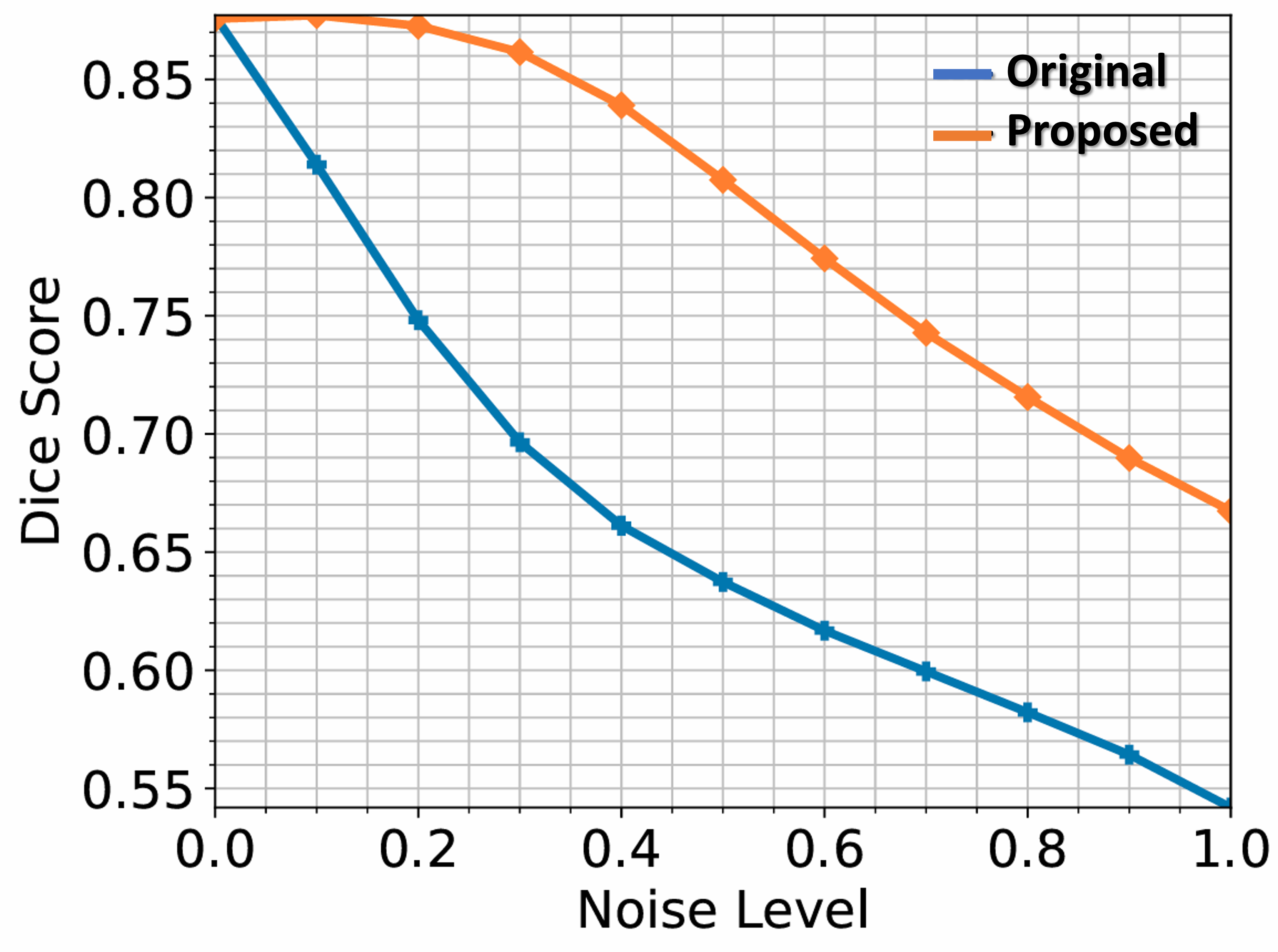} 
    \caption{Segmentation robustness vs.\ noise: averaged Dice for FHPS dataset with and without the proposed preprocessing.}
    \label{fig:dice_score_vs_noise}
\end{figure}
\noindent\textbf{Impact on segmentation.} On PSFHS, Figure~\ref{fig:dice_score_vs_noise} shows that applying the proposed method as a preprocessing step consistently increases combined Dice for both fetal head and pubic symphysis across all noise levels, with the largest improvements under severe noise. This indicates that the model not only enhances perceptual quality but also preserves anatomically meaningful edges required by downstream networks. Together with the noise-ladder results (Figs.~\ref{fig:new_gaussian_psnr}--\ref{fig:new_speckle_ssim}), these findings confirm robustness to \emph{combined} blur and noise and support the method’s plug-and-play use in clinical pipelines.

\section{CONCLUSION}
We presented proposed method, a blind, self-supervised framework for ultrasound enhancement that unifies deconvolution and speckle suppression without requiring clean targets, explicit PSF calibration, or noise variance estimation. The approach couples a Swin-based encoder–decoder with a generalized, physics-guided degradation model that exposes the network to diverse and realistic corruption modes drawn from actual ultrasound data.

Comprehensive experiments demonstrate consistent performance gain under Gaussian and speckle ladders across multiple datasets. Proposed method remains the top PSNR/SSIM performer with the advantage increasing in the most challenging regimes. Second, controlled blur studies show resolution recovery with narrower FWHM and higher peak gradients, while preserving anatomical boundaries.  Limitations include potential bias introduced by using NLLR to approximate clean-like targets and the use of 2D B-mode images. Future work will investigate end-to-end joint learning with segmentation networks, extension to 3D/temporal US, explicit uncertainty modeling, and lightweight variants for real-time deployment on scanners. 

\section*{Availability of Code}
The implementation of our proposed method is available on the following GitHub repository: \url{https://github.com/5y3datif/Blind_Ultrasound_Enhancement}.

\section*{ACKNOWLEDGMENT}
Shujaat Khan acknowledges the support from the King Fahd University of Petroleum \& Minerals (KFUPM) under Early Career Research Grant no. EC241027.

\bibliographystyle{IEEEtran}
\bibliography{References}

@article{hendrycks2016gaussian,
  title={Gaussian error linear units (gelus)},
  author={Hendrycks, Dan and Gimpel, Kevin},
  journal={arXiv preprint arXiv:1606.08415},
  year={2016}
}

@inproceedings{liu2021swin,
  title={Swin transformer: Hierarchical vision transformer using shifted windows},
  author={Liu, Ze and Lin, Yutong and Cao, Yue and Hu, Han and Wei, Yixuan and Zhang, Zheng and Lin, Stephen and Guo, Baining},
  booktitle={Proceedings of the IEEE/CVF international conference on computer vision},
  pages={10012--10022},
  year={2021}
}

@article{zhang2023practical,
  title={Practical blind image denoising via Swin-Conv-UNet and data synthesis},
  author={Zhang, Kai and Li, Yawei and Liang, Jingyun and Cao, Jiezhang and Zhang, Yulun and Tang, Hao and Fan, Deng-Ping and Timofte, Radu and Gool, Luc Van},
  journal={Machine Intelligence Research},
  volume={20},
  number={6},
  pages={822--836},
  year={2023},
  publisher={Springer}
}

@inproceedings{khan2019deep,
  title={Deep learning-based universal beamformer for ultrasound imaging},
  author={Khan, Shujaat and Huh, Jaeyoung and Ye, Jong Chul},
  booktitle={International Conference on Medical Image Computing and Computer-Assisted Intervention},
  pages={619--627},
  year={2019},
  organization={Springer International Publishing Cham}
}

@article{khan2020adaptive,
  title={Adaptive and Compressive Beamforming using Deep Learning for Medical Ultrasound},
  author={Khan, Shujaat and Huh, Jaeyoung and Ye, Jong Chul},
  journal={IEEE Transactions on Ultrasonics, Ferroelectrics, and Frequency Control},
  pages={1--1},
  year={2020},
  publisher={IEEE}
}

@inproceedings{khan2020unsupervised,
  title={Unsupervised Deconvolution Neural Network for High Quality Ultrasound Imaging},
  author={Khan, Shujaat and Huh, Jaeyoung and Ye, Jong Chul},
  booktitle={2020 IEEE International Ultrasonics Symposium (IUS)},
  pages={1--4},
  year={2020},
  organization={IEEE}
}

@article{khan2021variational,
  title={Variational Formulation of Unsupervised Deep Learning for Ultrasound Image Artifact Removal},
  author={Khan, Shujaat and Huh, Jaeyoung and Ye, Jong Chul},
  journal={IEEE Transactions on Ultrasonics, Ferroelectrics, and Frequency Control},
  volume={68},
  number={6},
  pages={2086--2100},
  year={2021},
  publisher={IEEE}
}

@article{khan2021switchable,
  title={Switchable and Tunable Deep Beamformer Using Adaptive Instance Normalization for Medical Ultrasound},
  author={Khan, Shujaat and Huh, Jaeyoung and Ye, Jong Chul},
  journal={IEEE Transactions on Medical Imaging},
  volume={41},
  number={2},
  pages={266--278},
  year={2021},
  publisher={IEEE}
}

@article{yu2002srad,
  title={Speckle reducing anisotropic diffusion},
  author={Yu, Yongjian and Acton, Scott T},
  journal={IEEE Transactions on image processing},
  volume={11},
  number={11},
  pages={1260--1270},
  year={2002},
  publisher={IEEE}
}

@article{bai2007fractional,
  title={Fractional-order anisotropic diffusion for image denoising},
  author={Bai, Jian and Feng, Xiang-Chu},
  journal={IEEE transactions on image processing},
  volume={16},
  number={10},
  pages={2492--2502},
  year={2007},
  publisher={IEEE}
}

@article{kollem2024fast,
  title={A fast computational technique based on a novel tangent sigmoid anisotropic diffusion function for image-denoising},
  author={Kollem, Sreedhar},
  journal={Soft Computing},
  volume={28},
  number={11},
  pages={7501--7526},
  year={2024},
  publisher={Springer}
}

@inproceedings{buades2005nonlocalmeans,
  title={A non-local algorithm for image denoising},
  author={Buades, Antoni and Coll, Bartomeu and Morel, J-M},
  booktitle={2005 IEEE computer society conference on computer vision and pattern recognition (CVPR'05)},
  volume={2},
  pages={60--65},
  year={2005},
  organization={Ieee}
}

@inproceedings{kervrann2007bayesian-nonlocalmeans,
  title={Bayesian non-local means filter, image redundancy and adaptive dictionaries for noise removal},
  author={Kervrann, Charles and Boulanger, J{\'e}r{\^o}me and Coup{\'e}, Pierrick},
  booktitle={International conference on scale space and variational methods in computer vision},
  pages={520--532},
  year={2007},
  organization={Springer}
}

@article{maleki2013anisotropic-nonlocalmeans,
  title={Anisotropic nonlocal means denoising},
  author={Maleki, Arian and Narayan, Manjari and Baraniuk, Richard G},
  journal={Applied and Computational Harmonic Analysis},
  volume={35},
  number={3},
  pages={452--482},
  year={2013},
  publisher={Elsevier}
}

@article{chang2000adaptive,
  title={Adaptive wavelet thresholding for image denoising and compression},
  author={Chang, S Grace and Yu, Bin and Vetterli, Martin},
  journal={IEEE transactions on image processing},
  volume={9},
  number={9},
  pages={1532--1546},
  year={2000},
  publisher={IEEE}
}

@article{liu2017efficient,
  title={Efficient single image dehazing and denoising: An efficient multi-scale correlated wavelet approach},
  author={Liu, Xin and Zhang, He and Cheung, Yiu-ming and You, Xinge and Tang, Yuan Yan},
  journal={Computer Vision and Image Understanding},
  volume={162},
  pages={23--33},
  year={2017},
  publisher={Elsevier}
}

@article{onur2022improved,
  title={Improved image denoising using wavelet edge detection based on Otsu’s thresholding},
  author={Onur, Tu{\u{g}}ba {\"O}zge},
  journal={Acta Polytechnica Hungarica},
  volume={19},
  number={2},
  pages={79--92},
  year={2022}
}

@article{zhang2022two,
  title={Two-step non-local means method for image denoising},
  author={Zhang, Xiaobo},
  journal={Multidimensional Systems and Signal Processing},
  volume={33},
  number={2},
  pages={341--366},
  year={2022},
  publisher={Springer}
}

@article{kong2024improved,
  title={An improved non-local means algorithm for CT image denoising},
  author={Kong, Huihua and Gao, Wenbo and Du, Xiaoshuang and Di, Yunxia},
  journal={Multimedia Systems},
  volume={30},
  number={2},
  pages={79},
  year={2024},
  publisher={Springer}
}

@article{rajwade2012image,
  title={Image denoising using the higher order singular value decomposition},
  author={Rajwade, Ajit and Rangarajan, Anand and Banerjee, Arunava},
  journal={IEEE Transactions on Pattern Analysis and Machine Intelligence},
  volume={35},
  number={4},
  pages={849--862},
  year={2012},
  publisher={IEEE}
}

@inproceedings{zhu2017non,
  title={A non-local low-rank framework for ultrasound speckle reduction},
  author={Zhu, Lei and Fu, Chi-Wing and Brown, Michael S and Heng, Pheng-Ann},
  booktitle={Proceedings of the IEEE conference on computer vision and pattern recognition},
  pages={5650--5658},
  year={2017}
}

@inproceedings{zhang2024fast,
  title={Fast and accurate estimation of low-rank matrices from noisy measurements via preconditioned non-convex gradient descent},
  author={Zhang, Jialun and Zhang, Richard Y and Chiu, Hong-Ming},
  booktitle={International Conference on Artificial Intelligence and Statistics},
  pages={3772--3780},
  year={2024},
  organization={PMLR}
}

@article{zhang2017beyond,
  title={Beyond a gaussian denoiser: Residual learning of deep cnn for image denoising},
  author={Zhang, Kai and Zuo, Wangmeng and Chen, Yunjin and Meng, Deyu and Zhang, Lei},
  journal={IEEE transactions on image processing},
  volume={26},
  number={7},
  pages={3142--3155},
  year={2017},
  publisher={IEEE}
}

@inproceedings{ronneberger2015unet,
  title={U-net: Convolutional networks for biomedical image segmentation},
  author={Ronneberger, Olaf and Fischer, Philipp and Brox, Thomas},
  booktitle={Medical image computing and computer-assisted intervention--MICCAI 2015: 18th international conference, Munich, Germany, October 5-9, 2015, proceedings, part III 18},
  pages={234--241},
  year={2015},
  organization={Springer}
}

@inproceedings{krull2019noise2void,
  title={Noise2void-learning denoising from single noisy images},
  author={Krull, Alexander and Buchholz, Tim-Oliver and Jug, Florian},
  booktitle={Proceedings of the IEEE/CVF conference on computer vision and pattern recognition},
  pages={2129--2137},
  year={2019}
}

@inproceedings{batson2019noise2self,
  title={Noise2self: Blind denoising by self-supervision},
  author={Batson, Joshua and Royer, Loic},
  booktitle={International conference on machine learning},
  pages={524--533},
  year={2019},
  organization={PMLR}
}

@inproceedings{gondara2016medical,
  title={Medical image denoising using convolutional denoising autoencoders},
  author={Gondara, Lovedeep},
  booktitle={2016 IEEE 16th international conference on data mining workshops (ICDMW)},
  pages={241--246},
  year={2016},
  organization={IEEE}
}

@inproceedings{nanthini2024dl,
  title={A DL-DeNoiseNet Framework for Robust Image Denoising and Artifact Reduction in Medical Image Diagnostics},
  author={Nanthini, S and Maindola, Meenakshi and Madhu, R and Ravishankar, MD and Gandhi, Jay and Patil, Harshal},
  booktitle={2024 International Conference on Integrated Intelligence and Communication Systems (ICIICS)},
  pages={1--7},
  year={2024},
  organization={IEEE}
}

@article{seoni2024all,
  title={All you need is data preparation: A systematic review of image harmonization techniques in Multi-center/device studies for medical support systems},
  author={Seoni, Silvia and Shahini, Alen and Meiburger, Kristen M and Marzola, Francesco and Rotunno, Giulia and Acharya, U Rajendra and Molinari, Filippo and Salvi, Massimo},
  journal={Computer Methods and Programs in Biomedicine},
  pages={108200},
  year={2024},
  publisher={Elsevier}
}

@article{yap2018udaitb,
  author={Yap, Moi Hoon and Pons, Gerard and Martí, Joan and Ganau, Sergi and Sentís, Melcior and Zwiggelaar, Reyer and Davison, Adrian K. and Martí, Robert},
  journal={IEEE Journal of Biomedical and Health Informatics}, 
  title={Automated Breast Ultrasound Lesions Detection Using Convolutional Neural Networks}, 
  year={2018},
  volume={22},
  number={4},
  pages={1218-1226},
  doi={10.1109/JBHI.2017.2731873}
}

@article{lu2022jnu,
  title={The JNU-IFM dataset for segmenting pubic symphysis-fetal head},
  author={Lu, Yaosheng and Zhou, Mengqiang and Zhi, Dengjiang and Zhou, Minghong and Jiang, Xiaosong and Qiu, Ruiyu and Ou, Zhanhong and Wang, Huijin and Qiu, Di and Zhong, Mei and others},
  journal={Data in brief},
  volume={41},
  pages={107904},
  year={2022},
  publisher={Elsevier}
}

@inproceedings{xia2017xpie,
  title={What is and what is not a salient object? learning salient object detector by ensembling linear exemplar regressors},
  author={Xia, Changqun and Li, Jia and Chen, Xiaowu and Zheng, Anlin and Zhang, Yu},
  booktitle={Proceedings of the IEEE conference on computer vision and pattern recognition},
  pages={4142--4150},
  year={2017}
}

@article{chen2024psfhs,
  title={PSFHS: intrapartum ultrasound image dataset for AI-based segmentation of pubic symphysis and fetal head},
  author={Chen, Gaowen and Bai, Jieyun and Ou, Zhanhong and Lu, Yaosheng and Wang, Huijin},
  journal={Scientific Data},
  volume={11},
  number={1},
  pages={436},
  year={2024},
  publisher={Nature Publishing Group UK London}
}

@article{guo2023blind,
  title={Blind image despeckling using a multiscale attention-guided neural network},
  author={Guo, Yu and Lu, Yuxu and Liu, Ryan Wen and Zhu, Fenghua},
  journal={IEEE Transactions on Artificial Intelligence},
  volume={5},
  number={1},
  pages={205--216},
  year={2023},
  publisher={IEEE}
}

@inproceedings{zamir2022restormer,
  title={Restormer: Efficient transformer for high-resolution image restoration},
  author={Zamir, Syed Waqas and Arora, Aditya and Khan, Salman and Hayat, Munawar and Khan, Fahad Shahbaz and Yang, Ming-Hsuan},
  booktitle={Proceedings of the IEEE/CVF conference on computer vision and pattern recognition},
  pages={5728--5739},
  year={2022}
}

\begin{IEEEbiography}[{\includegraphics[width=1in,height=1.25in,clip,keepaspectratio]{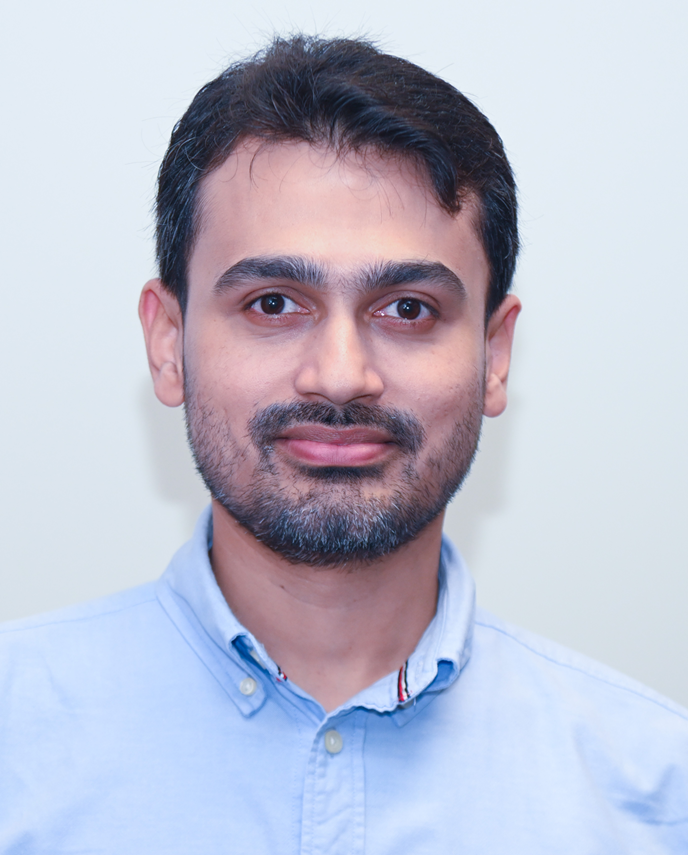}}]{Shujaat Khan } is an Assistant Professor in the Department of Computer Engineering and a Fellow at the Saudi Data and AI Authority (SDAIA) and King Fahd University of Petroleum \& Minerals (KFUPM) under the SDAIA-KFUPM Joint Research Center for Artificial Intelligence (JRC-AI) at KFUPM, Dhahran, KSA. Prior to joining KFUPM, he was a Senior AI Scientist at Digital Technology \& Innovation, Siemens Medical Solutions USA, Inc. He earned his Ph.D. from the Department of Bio and Brain Engineering at the Korea Advanced Institute of Science and Technology (KAIST), Daejeon, South Korea, in 2022. He was a researcher with Synergistic Bioinformatics (SynBi) and the Bio Imaging, Signal Processing Learning (BISPL) Labs at KAIST. His research interests include machine learning, optimization, inverse problems, and signal processing, with a focus on biomedical and bioinformatics applications. He has authored works in \textit{Medical Image Analysis}, \textit{IEEE Transactions on Medical Imaging}, \textit{IEEE Transactions on Computational Imaging}, and \textit{IEEE Transactions on Ultrasonics, Ferroelectrics, and Frequency Control}, and is an inventor on several ultrasound imaging patents.
\end{IEEEbiography}

\begin{IEEEbiography}[{\includegraphics[width=1in,height=1.25in,clip,keepaspectratio]{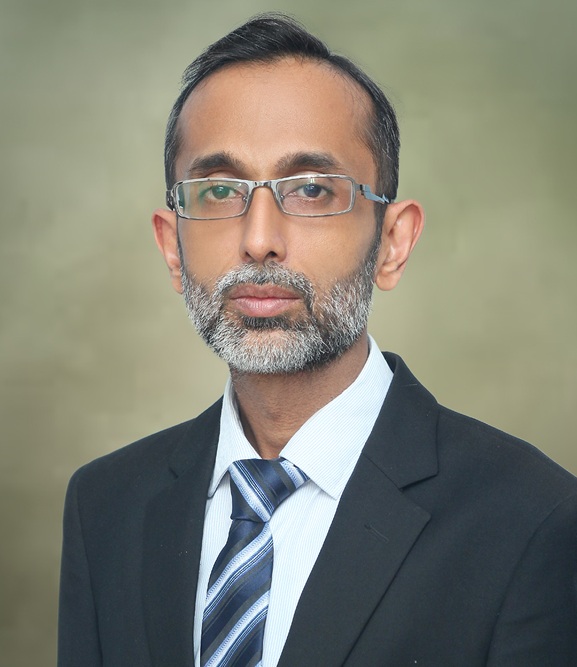}}]{Syed Muhammad Atif } is an Assistant Professor in the Department of Computer Science \& Information Technology, Sir Syed University of Engineering \& Technology (SSUET), Karachi, Pakistan. Prior to joining SSUET, he was an Assistant Professor in the Department of Computer Science \& Software Engineering, Ziauddin University, Karachi, Pakistan. He earned his Ph.D. from the Graduate School of Science \& Engineering at the Karachi Institute of Economic \& Technology (KIET), Karachi, Pakistan, in 2022. His research interests include machine learning, matrix factorization, and network tomography. 
\end{IEEEbiography}

\begin{IEEEbiography}[{\includegraphics[width=1in,height=1.25in,clip,keepaspectratio]{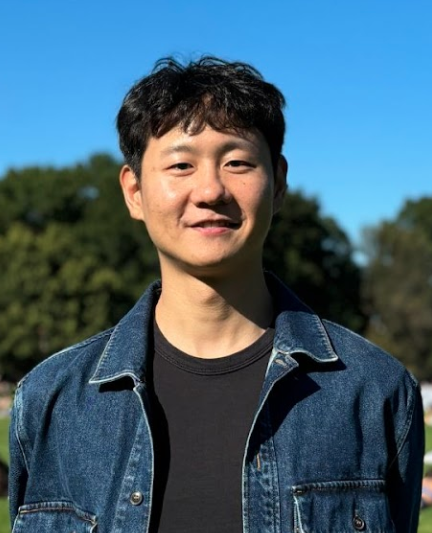}}]{Jaeyoung Huh } is a Postdoctoral Researcher at Siemens Healthineers, Princeton, NJ, USA, working on ultrasound image–guided robotic manipulation. He received the B.S. degree in Electronic Engineering from Hanyang University, Seoul, South Korea, in 2017, and the M.S. and Ph.D. degrees in Bio and Brain Engineering from KAIST, Daejeon, South Korea, in 2020 and 2024, respectively, where he was with the Bio Imaging, Signal Processing \& Learning (BISPL) Lab under Prof.\ Jong Chul Ye. His research interests include deep learning, computer vision, and medical/ultrasound imaging. He has authored works in \textit{Medical Image Analysis}, \textit{IEEE Transactions on Medical Imaging}, \textit{IEEE Transactions on Computational Imaging}, and \textit{IEEE Transactions on Ultrasonics, Ferroelectrics, and Frequency Control}, and is an inventor on several ultrasound imaging patents.
\end{IEEEbiography}

\begin{IEEEbiography}[{\includegraphics[width=1in,height=1.25in,clip,keepaspectratio]{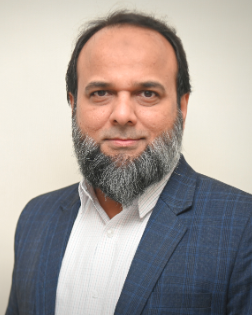}}]{Syed Saad Azhar Ali } is the Acting Chairman and an Assistant Professor with the Department of Aerospace Engineering, King Fahd University of Petroleum \& Minerals (KFUPM), Dhahran, Saudi Arabia. He is also a Research Scholar with the Interdisciplinary Research Center for Smart Mobility and Logistics at KFUPM. He received the M.S. (2002) and Ph.D. (2007) degrees in Electrical Engineering from KFUPM. Previously, he held faculty appointments at Universiti Teknologi PETRONAS (Assistant Professor, 2014–2017; Associate Professor, 2018–2022), Iqra University (Associate Professor, 2009–2013), and Air University Islamabad (Assistant Professor, 2007–2009). His research interests include machine learning for healthcare and neuroimaging (EEG and MRI), explainable AI, human mobility and gait analysis, UAV modeling and control, and intelligent transportation systems.
\end{IEEEbiography}

\vfill\pagebreak

\end{document}